# Triboelectrification of a dense metal-organic framework for resilient mechanical energy harvesters

Chuzhan Jin, Debayan Mondal, Jiahao Ye, Jin-Chong Tan*
Multifunctional Materials & Composites (MMC) Laboratory, Department of Engineering Science, University of Oxford, United Kingdom.

**Corresponding author*: jin-chong.tan@eng.ox.ac.uk

**Abstract**

Triboelectric nanogenerators (TENGs) incorporating metal-organic frameworks (MOFs) have largely been designed around porous architectures, based on the assumption that high internal surface area is the primary driver for triboelectric enhancement. Herein, we demonstrate that a dense, nominally nonporous MOF called ZIF-zni can instead function as an effective high-loading filler within a tribopositive polyurethane (PU) matrix, offering a design strategy for harnessing interfacial electromechanical effects. A 20 wt% ZIF-zni@PU composite delivers an output voltage of 470±15 V and a peak power density of 1.31±0.03 W $m^{-2}$ against polydimethylsiloxane (PDMS). The device exhibits stable performance over ~94,000 cycles under a contact force of ~100 N, and it remains operational under higher impact beyond 500 N. This concept enables the demonstration of a proof-of-concept triboelectric floor tile. Combined experimental and theoretical studies indicate that the performance enhancement arises from favourable interfacial polarization, reduced dielectric screening, and surface accessible ZIF-zni rich domains, rather than porosity alone. These features are accompanied by reduced work of adhesion and modified surface roughness, hence improving contact electrification. These findings establish dense MOFs as an effective triboelectric filler and identify interfacial electronic structure and polarization as key design parameters for engineering mechanical energy harvesters and self-powered sensors.

**Keywords:** Triboelectric nanogenerator; metal-organic framework; zeolitic imidazolate framework; dielectric properties; mechanical energy; nearfield nanospectroscopy; Kelvin probe force microscopy; density functional theory; mechanical properties.

## Introduction

The escalating global energy demand has intensified the search for sustainable and alternative power sources, for instance through the scavenging of waste mechanical energy omnipresent in our surroundings[1]. Triboelectric nanogenerators (TENGs) have emerged as a compelling solution, converting low-frequency mechanical motions into electrical power through coupled contact electrification and electrostatic induction[2-5].Their inherent advantages, including lightweight construction, broad material versatility, and tolerance to irregular mechanical oscillations, make TENGs potentially useful for powering small devices encompassing wearable sensors, health monitoring devices, and compact electronics[6-8]. However, the transition from laboratory demonstrations to practical deployment critically depends on precise engineering of durable triboelectric materials, with particular emphasis on mechanically resilient polymer-based composites that constitute the functional active layers.

Metal-organic frameworks (MOFs) have emerged as an attractive functional nanofiller owing to their long-range crystalline order and vastly tunable physical and chemical properties[9-12]. Early efforts integrating MOFs into tribonegative polymers such as polydimethylsiloxane (PDMS) demonstrated appreciable enhancements in triboelectric output[13]. For example, Wen *et al.* reported that incorporation of 4 wt% ZIF-8 in PDMS generated an output voltage of 176 V with a nominal sample area of 2 × 2 $cm^2$ and a current density of ~4 $\mu A\ cm^{-2}$ [14], while UiO-66-$NH_2$ based PDMS composites delivered power densities up to ~1.7 W $m^{-2}$ (3 × 3 $cm^2$ nominal contact area), attributed to its electron withdrawing group[15]. More intriguingly, Ye *et al.* showed that only a 1 wt% MOF filler of the comparatively dense, halogenated ZIF-72 in PDMS (henceforth, denoted as ZIF-72@PDMS) yielded output voltages approaching 578 V for a 3 × 3 $cm^2$ device, along with a power density of ~5 W $m^{-2}$, surpassing many highly porous MOFs incorporating a similar level of filler loading[16]. This observation suggests that structural density and dielectric properties may supersede the effect of internal porosity to yield enhanced triboelectric output.

Despite above recent developments, the field has largely focused on porous MOFs dispersed within tribonegative polymer matrices. This prevailing strategy overlooks a key materials principle since most imidazolate- and carboxylate-based MOFs are inherently tribopositive, and therefore they have the propensity to donate electrons during contact electrification. When combined with a tribonegative polymer such as PDMS, the opposing charge-transfer tendencies create a polarity mismatch, which inherently limits net charge accumulation and transfer. Hitherto, a more rational design strategy, namely by pairing tribopositive MOFs with tribopositive polymers, has received comparatively little systematic investigation. Limited work with polyacrylonitrile (PAN) or polyvinylpyrrolidone (PVP) composites has shown promise, with ZIF-8@PAN nanofibers achieving power densities of ~1.9 W $m^{-2}$ and MIL-88A@PVP/PAA films producing output voltages of ~54 V and current density of ~84 nA $cm^{-2}$ for the 2 × 2 $cm^2$ sample [17,18]. However, these reports rely predominantly on electrospinning, which yields high surface area morphologies while limiting achievable filler content loadings (wt%) and precluding a systematic exploration of dense, nominally nonporous MOFs at practically relevant mass fractions.

In this work, we address these gaps by selecting the ZIF-zni, the densest known zeolitic imidazolate framework (ZIF) structure[19], as a model dense system, nominally nonporous and tribopositive filler for examining triboelectric enhancement in a low-porosity MOF@polymer composite film. Unlike conventionally porous ZIFs, ZIF-zni (topology is zni) has a very low reported solvent-accessible volume ($SAV \approx 1.5\%$), thereby limiting the expected contribution of internal porosity and enabling exploration of possible dense-framework effects under a controlled composite design[19]. First-principles calculations employing *ab initio* density functional theory (DFT) were used to identify basic electronic structures and vibrational characteristics of ZIF-zni that may influence interfacial polarization in TENG. We hypothesized that the nonporous topology yields high-Born effective charges, enabling direct strain-to-polarization conversion along with the band-edge asymmetry that spatially separates charge carriers to suppress subsequent charge recombination. Herein, we determined a static dielectric constant, $\kappa$ or $\varepsilon'$ value of ~2, which is significantly higher than other porous ZIFs[20]. Dense framework limits dipole reorientation and weakens dynamic electrostatic screening at the interface. These effects may contribute alongside interfacial morphology and polymer-filler interactions, whereby direct comparisons with porous analogues are necessary to further isolate the role of framework density.

Subsequently, guided by the aforementioned considerations, ZIF-zni was incorporated into a tribopositive polyurethane (PU) matrix to form composite films with systematically varied filler contents ranging from 2 to 25 wt% *via* doctor blade coating, representing a scalable and low-cost method capable of producing uniform composite films with a high filler content.

Kelvin probe force microscopy (KPFM) was used to evaluate relative surface-potential charges, while nanoindentation and near-field infrared nanospectroscopy (nano-FTIR) provided insights into their local mechanical and vibrational properties. The resulting TENG devices, operated in contact-separation mode, were evaluated for long-term stability, humidity-dependent response (nominal 10-85% RH) and high impact load (~610 N) including operation within a practical TENG floor tile prototype (20 cm × 20 cm) to assess their practical use under a more realistic operating condition. By emphasizing dense, nominally nonporous MOFs as functional fillers, this study shows that descriptors beyond internal porosity of open frameworks should be considered when designing MOF-based triboelectric composites.

## Results and discussion

### Synthesis and characterization of ZIF-zni@PU polymer composites

The structural and vibrational characterization was performed to confirm that ZIF-zni retains its crystalline framework after incorporation into the polyurethane (PU) matrix. The preparation of ZIF-zni@PU composites is illustrated in Figure 1a. The $\alpha$-phase of ZIF-zni was synthesised through a scalable one-pot method using deionised water as a green solvent[21]. Powder X-ray diffraction (PXRD) confirmed the phase purity, with the experimental pattern matching the simulated diffraction pattern (Figure 1a). The SEM image reveals micron-sized, rod-like ZIF-zni crystals with an average length of approximately 1.6 μm (Figure 1a). The vibrational spectra of the as-synthesised ZIF-zni crystals were consistent with the simulated infrared (IR) and Raman spectra determined from DFT calculations, confirming the assignment of characteristic vibrational modes in the framework. The far-infrared (FIR) spectrum (Figure 1b(i)) revealed the collective terahertz (THz) vibrational modes of ZIF-zni. Weak bands below 200 $cm^{-1}$ are attributed to lattice vibrations (phonons), while bands in the 200-350 $cm^{-1}$ region correspond to the Zn-N stretching and imidazolate ring deformation. A prominent band at approximately 650-700 $cm^{-1}$ was assigned to the out-of-plane bending of the imidazolate linkers. Although the experimental spectrum exhibited broader features than the DFT prediction, likely owing to structural disorder and anharmonic effects, the calculated and experimental spectra are in close agreement in the positions of the principal vibrational modes. The corresponding Fourier-transform infrared spectroscopy (ATR-FTIR) spectrum (Figure 1b(ii)) displayed a characteristic absorption band at approximately 1495 $cm^{-1}$, assigned to the C=N and C=C stretching modes of the imidazolate linkers. In the low-frequency range of Raman vibrations using the Eclipse filter (Figure 1c(i)), the experimental spectrum was dominated by weak and broad features below ~200 $cm^{-1}$ ($\lesssim$ 6 THz), which were attributed to lattice vibrations and Zn-N tetrahedral modes. The peaks around ~1185 $cm^{-1}$ and ~1294 $cm^{-1}$ in the Raman spectrum were both associated with imidazolate ring vibrations (Figure 1c(ii)). The peak around ~1490 $cm^{-1}$ had a small shift in wavenumber, since it involved C-H bending coupled with ring deformation of the imidazolate linker which was more sensitive to the local chemical environment and its structural rigidity.

The ZIF-zni@PU slurry was cast using a doctor blade to obtain uniform composite films, followed by oven drying. The resulting films were characterized by attenuated total reflectance ATR-FTIR and PXRD. As shown in Figure 1d, the ATR-FTIR spectra of the composite films exhibited characteristic vibrational bands originating from both PU and ZIF-zni. The urethane carbonyl stretching band of PU appeared at 1700-1720 $cm^{-1}$ and remained observable across all compositions, indicating preservation of the polymer structure after incorporation of ZIF-zni. The spectral region between 1500 and 1590 $cm^{-1}$ contained contributions from the imidazolate C=N/C=C stretching modes of ZIF-zni together with the amide II band of PU. In addition, two characteristic ZIF-zni bands located at approximately 667 and 950 $cm^{-1}$, assigned to the out-of-plane imidazolate ring bending and ring vibrational modes, respectively, became progressively more pronounced with an increasing wt% loading of ZIF-zni. The simultaneous retention of the characteristic PU bands and emergence of ZIF-zni vibrational features confirms successful incorporation of ZIF-zni into the PU matrix. The crystalline structure of ZIF-zni within the composite films was further verified by PXRD (Figure 1e). The synthesized ZIF-zni exhibited the characteristic diffraction peak corresponding to the (400) plane at $2\theta \approx 15°$. This reflection was retained in all the composite films, indicating preservation of the ZIF-zni crystalline phase after incorporation into the PU matrix. Its gradual intensity increase with ZIF-zni loading is consistent with higher filler content, although we note that film thickness and sample orientation may also influence peak intensity (Table S1, Supporting Information (SI)).

**Triboelectric performance of ZIF-zni@PU composites**

The ZIF-zni@PU TENG device was fabricated for a contact-separation test configuration against pristine PDMS (Figure 2a) and operated under a cyclic motion. Based on the fundamental working mechanism of a TENG, the ZIF-zni@PU against PDMS device operates in a dielectric-to-dielectric contact-separation mode, in which ZIF-zni@PU serves as the tribopositive layer and PDMS serves as the tribonegative layer. The proposed mechanism is elucidated as follows. Initially, the two triboelectric layers on opposing electrodes are retracted far apart, thus no electron transfer occurs (see SI, Figure S1(a)i). When an external force is applied, the tribopositive ZIF-zni@PU layer comes into contact with the tribonegative PDMS layer. As a result of triboelectrification, equal amounts of positive and negative charges are generated on the opposing contacting surfaces of the two materials (Figure S1(a)ii). After the external force is released, the separation of the two layers creates a potential difference that drives electrons to flow through the external circuit *via* the ITO-coated PET electrodes as illustrated in Figure S1(a)iii. When the layers are fully separated, electrostatic equilibrium is reached (Figure S1(a)iv). During the subsequent pressing process, the layers move toward each other again, causing electrons to flow in the opposite direction through the external circuit (Figure S1(a)v). This contact–separation cycle is repeatedly driven by periodic mechanical motion, generating a continuous alternating electrical output.

ZIF-zni was incorporated into PU at 2, 5, 10, 15, 20 and 25 wt% to obtain ZIF-zni@PU composite films used as the tribopositive layer; the corresponding microscopic images and photos of the fabricated devices are shown in Figure S2a-b. For the wt.% filler-loading-dependent measurements shown in Figure 2b–d, the maximum impact force was maintained at ~100 N and the maximum separation distance was fixed at 2.5 mm. The impact force was monitored using a load cell (Figure S3a), and the 2.5 mm separation distance was selected based on the distance-dependent output test (Figure S3b). The open-circuit voltage of pristine PU reached 95 ± 12 V, whereas incorporation of ZIF-zni enhanced the output, with a maximum output of 470 ± 15 V at 20 wt% of ZIF-zni, approximately 5 times higher than the pristine PU (Fig. 2b, Figure S4a). In contrast to many reported porous tribopositive MOF@polymer composites, which often show optimal output at a relatively lower filler loading of 1-10 wt%[13,22-24], we found the ZIF-zni@PU system retains effective charge generation at relatively higher filler contents. These findings suggest that using dense MOF fillers can be an effective design strategy for triboelectric composites. It can be seen in Figures 2c-d that the transferred charge and short-circuit current of the 20 wt% device increased to 21.5 ± 1.5 μC $m^{-2}$ and 10.3 ± 0.6 mA, respectively (Figure S4b), mirroring the current density trend (Figure S4c), while the peak-to-peak voltage reached 773 ± 16 V. Long-term cyclic tests showed an approximately stable output up to ~94,000 cycles, followed by a gradual decrease until 187,000 cycles, which may arise from multiple factors including variation in interfacial contact and material transfer, viscoelastic relaxation, environmental fluctuations, or slight mechanical misalignment during extended operation (Fig. 2e, Figure S4d).

The device achieved a maximum peak power density of 1.31 ± 0.03 W $m^{-2}$ at an optimal load resistance of 30 MΩ (Fig. 2f). Furthermore, we show that the alternating current generated by the ZIF-zni@PU TENG could be rectified and stored in capacitors, which were rapidly charged by high voltages under 2 Hz operation; for instance, a 0.1 μF capacitor reached approximately 12 V within 40 s (Fig. 2g). Finally, we demonstrate that the rectified output was sufficient to directly illuminate 240 commercial LEDs during periodic contact-separation operation (Fig. 2h), demonstrating the potential of the ZIF-zni@PU TENG for driving low-power demonstrations under laboratory conditions.

**Performance analysis and underlying mechanisms of ZIF-zni TENG device**

To elucidate the origin of the enhanced triboelectric output, Kelvin probe force microscopy (KPFM) was performed on pristine PU and ZIF-zni@PU composite films (Figure 3a). The surface potential increased progressively from 1.11 V for pristine PU to 1.37 V for the 20 wt% composite, indicating a systematic modification of the surface potential upon incorporation of ZIF-zni. Based on the measured contact potential difference ($V_{\mathrm{CPD}}$), the corresponding work function was calculated according to Equation (1):

$$\Phi_{\mathrm{sample}}\ (\mathrm{eV}) = \Phi_{\mathrm{tip}}\ (\mathrm{eV}) - V_{\mathrm{CPD}}\ (\mathrm{V}) \qquad (1)$$

where $\Phi_{\mathrm{sample}}$ is the work function of the ZIF-zni@PU composite, $\Phi_{\mathrm{tip}}$ is the work function of the Pt/Ir-coated AFM tip used herein with a reported value of 4.6 eV, and $V_{\mathrm{CPD}}$ is the measured contact potential difference. Because no reference-standard calibration was performed, the work-function values should be interpreted as relative estimates rather than absolute values.

Based on the measured $V_{\mathrm{CPD}}$ values and this tip reference, the work function decreased from 3.49 eV for pristine PU to 3.34, 3.29, and 3.23 eV for the 10, 15, and 20 wt% ZIF-zni@PU composites, respectively. As illustrated in Figure S5, this trend reveals a relative shift in the surface electronic state of the composite. The shift is consistent with an enhanced electron-donating tendency of the ZIF-zni@PU surface, but it should not be interpreted as a calibrated absolute work-function alignment with PDMS.

Pull-off force measurements were performed using a flat punch nanoindenter equipped with a cylindrical punch, with a nominal diameter of 10 μm (Figure S6a)[25]. The pull-off force declined progressively with increasing ZIF-zni wt% loading, indicating weakened interfacial adhesion between the punch and the composite surface. The work of adhesion was subsequently determined by integrating the area under the unloading segment of the load versus depth curve (Figure S6b-h). As shown in Figure 3b, the work of adhesion exhibited a similar trend, decreasing with increasing filler content. We reasoned that, as the filler wt.% increases the mechanical stiffness of the composite surface rises, thus leading to a fall in surface adhesion. A pronounced reduction was observed up to 10 wt%, followed by a gradual decrease toward a plateau at higher loadings. The adhesion work decreased from 24.7 pJ for the pristine sample to 6.3 pJ at 20 wt%, with only a slight increase at 25 wt%, which may be attributed to local filler aggregation. The consistent reductions in both the pull-off force and adhesion work indicate a weakened interfacial interaction between the flat punch and the composite surface, facilitating interfacial detachment during separation. Reduced adhesion could lead to more efficient contact separation during cyclic operation.

Besides interfacial adhesion, the dielectric response of the triboelectric layer can also influence the electrical output. As shown in Figure 3c and Table S2 (SI), the dielectric constant increased up to 20 wt%, followed by a slight decrease at 25 wt%. The dielectric constant of pristine ZIF-zni was determined to be $\varepsilon'$ or $\kappa = 2.16 \pm 0.07$ from pellet measurements using a parallel-plate configuration. Incorporation of ZIF-zni into the PU matrix therefore modified the dielectric response of the composite films, contributing to the observed electrical-output trend.

To investigate the contribution of surface morphology, the three-dimensional surface topography of the composite films was characterized (Figure 3d). Representative measurements showed the rms roughness values ($S_q$) increasing from 11.5 μm for the 2 wt% composite to 28.1 μm for the 25 wt% composite, suggesting a greater surface heterogeneity with increasing filler loading. The increased roughness is expected to modify the real contact area and local pressure distribution, which can affect triboelectric charge generation. However, despite exhibiting the highest roughness, the 25 wt% composite produced a lower electrical output than the 20 wt% sample. This result suggests that surface roughness alone is insufficient to determine device performance. Instead, the enhanced triboelectric output is associated with a balance among reduced work of adhesion, increased dielectric response, and modified interfacial contact geometry. Because the cured film thickness increased slightly with ZIF-zni loading (Table S1), thickness-related changes in capacitance and contact mechanics may also contribute to the electrical-output trend. The lower output at 25 wt% may reflect filler agglomeration, increased thickness, or surface defects that offset the benefit of higher roughness.

Apart from the macroscopic investigation, nano-FTIR was employed to investigate the nanoscale chemical distribution at the subsurface of the ZIF-zni@PU composite. As shown in Figure 4a, ZIF-zni crystals were identified on the PU surface, while the corresponding near-field topography (Figure 4b) revealed elevated crystalline domains highlighted in yellow. A 12-point line scan across the selected region (Figure 4c), together with the IR intensity heatmap (Figure 4d), showed systematic variations in the characteristic ZIF-zni vibrational band at approximately 1500 $cm^{-1}$, assigned to the imidazolate linker. In contrast, pristine PU (Figure 1d) exhibited no IR absorption at this wavenumber but instead displayed the characteristic N-H in plane bending vibration at approximately 1530 $cm^{-1}$. Additional PU bands at approximately 1220 $cm^{-1}$ (C–N stretching) and 1720 $cm^{-1}$ (C=O stretching) are evident at scan positions 1-4 and 11-12 (Figure 4d). Weak carbonyl absorption around 1720 $cm^{-1}$ observed between scan positions 6 and 10 indicates the coexistence of ZIF-zni rich and PU rich regions on the composite surface, confirming that the ZIF-zni crystal surface is partially covered by a thin layer of PU. These observations indicate that ZIF-zni crystals are located at or close to the composite surface, where they may influence interfacial interactions during contact electrification while remaining mechanically integrated with the polymer matrix.

Nano-FTIR measurements were further performed on the PDMS tribonegative film following repeated contact-separation cycles (Figure S7a,b). Spectra collected from selected locations (scan 1 and 2 in Figure S7b,) exhibited characteristic ZIF-zni bands at approximately 1140, 1220 and 1300 $cm^{-1}$ together with PU bands at approximately 1530 and 1730 $cm^{-1}$. The simultaneous observation of both spectral signatures within the same probing volume indicated that material containing both ZIF-zni and PU was present on the PDMS surface after the cyclic test. This observation is consistent with material transfer or interfacial reconstruction during repeated contact-separation, resulting in overlapping infrared contributions from both components.

To better examine the local spatial distribution of ZIF-zni within the PU matrix, we performed pseudoheterodyne (PsHet) near-field nanoimaging at 1496 $cm^{-1}$, corresponding to the nearfield absorption band of ZIF-zni (Figure S7c). The optical amplitude O2A image (Figure 4e), together with the corresponding signals for O2P (Figure 4f), O3A (Figure 4g), O3P (Figure 4h), and normalized O3A/O2A and O3P-O2P images (Figure S7d,e), clearly distinguished nanoscale regions exhibiting strong infrared absorption from PU dominated areas. Since pristine PU exhibited negligible absorption at 1496 $cm^{-1}$ (Figure 1d), the observed contrast indicated that ZIF-zni was exposed at or located close to the composite surface. Such surface accessibility may increase interfacial interactions during contact-separation and contribute to the enhanced triboelectric performance observed for the composite. Collectively, the nano-FTIR and PsHet analyses indicated that the nanoscale spatial distribution of ZIF-zni at the composite surface is likely an important factor influencing interfacial contact electrification.

Raman spectral mapping was subsequently employed to evaluate the structural stability of the composite following prolonged contact-separation cycles (Figures S8 and S9). Fiducial markers enabled the same regions to be relocated before and after testing, allowing direct spatial comparison. For the 20 wt% ZIF-zni@PU composite, Raman maps clearly distinguished ZIF-zni crystals from the surrounding PU matrix (Figure S8a,b), while spectra collected after overnight cycling closely matched those obtained before testing and the reference spectrum of pristine ZIF-zni (Figure 1c). The average Raman spectrum of the composite was presented in Figure S8c. The principal peak positions of the PU carbonyl stretching band at approximately 1720 $cm^{-1}$ and the aromatic C=C stretching band at approximately 1610 $cm^{-1}$ were retained after testing, although intensity variations were observed. These results indicate that no clear peak-position evidence for chemical degradation of PU or ZIF-zni was detected during prolonged cycling. In contrast, PDMS exhibited moderate variations in Raman intensity at approximately 1280 and 1400 $cm^{-1}$ while retaining unchanged peak positions (Figure S9). These intensity variations were consistent with minor surface rearrangement or local morphological changes rather than chemical modification of the polymer. Taken together, the Raman results did not provide evidence of detectable chemical degradation of either the ZIF-zni@PU composite or the PDMS surface during prolonged contact-separation. Consequently, the gradual decrease in electrical output observed after ~94,000 cycles is more likely associated with changes in the interfacial contact condition, material transfer, or mechanical effects such as viscoelastic relaxation than with irreversible chemical degradation.

**TENG performance under high contact force impact**

Instead of conventional output measurements, a higher load test rig equipped with a permanent electromagnetic V400 shaker was employed to apply higher forces of up to approximately 610 N, corresponding to the representative adult stepping loads (Figure S1c). The 20 wt% ZIF-zni@PU TENG was first compared with a pristine PU TENG under a maximum force of ~230 N, where the ZIF-zni@PU device exhibited a more pronounced increase in output with increasing force, supporting the contribution from ZIF-zni filler (Figure S10). In both devices, particularly in the voltage output measurements (Figure S10a,e), the down peaks (separation) were consistently longer than the up peaks (contact), which can be attributed to asymmetric contact-separation dynamics arising from the intrinsic stickiness and viscoelasticity of PDMS. Rapid contact during loading produces sharp up peaks, whereas adhesive interactions and viscoelastic relaxation during unloading delay separation, resulting in broadened down peaks. When the applied force was increased to ~610 N by reducing the separation distance to 2 mm, this behaviour persisted, while the output reached ~550 V and ~53 µA with single peaks shown in the enlarged views (Figure 5a-b). To evaluate the device performance under a more realistic operating condition, a larger 20 cm × 20 cm TENG floor tile module was fabricated, comprising a 15 cm × 15 cm active contact area and a 0.5 cm thick polycarbonate support (Figure S11). The device was tested under human stepping at different rates, defined as fast (1 step $s^{-1}$), medium (0.5 step $s^{-1}$), and slow (0.12 step $s^{-1}$) (Figure 5c(ii–iii), Figure S12 and Movie S2 in SI). The output amplitude depended on body weight, walking pattern, and contact duration, with a representative approximately 61 kg individual generating stable and repeatable pulses of ~150-289 V per step or ~550-680 V peak-to-peak and ~34 µA or ~120 µA peak-to-peak at the medium stepping rate of 0.5 step $s^{-1}$ (Figure 5c(ii–iii)). The generated signals were rectified and stored in a 1 µF capacitor during repeated stepping by a ~60 kg participant at a medium walking pace, producing a stepwise increase in voltage that confirms electrical energy accumulation under human stepping (Figure 5c(iv)). The capacitor energy increment can be estimated using Equation (2):

$$\Delta E = \frac{1}{2} C \left(V_2^2 - V_1^2\right) \qquad (2)$$

where $C$ is the storage capacitance and $V_1$ and $V_2$ are the capacitor voltages before and after a charging interval. A rigorous per-step harvested energy calculation would require synchronous $V(t)$ and $I(t)$ measurements under the same external load, or step-resolved capacitor-voltage increments. Accordingly, Figure 5c(iv) is intended to demonstrate the capability for electrical energy storage rather than to quantify the harvested energy per step.

## Humidity-responsive triboelectric behaviour

The 20 wt% ZIF zni@PU TENG exhibits a distinct humidity dependent triboelectric response over the investigated RH range (Figure S13a). The output voltage remains nearly constant at low RH and gradually decreases with increasing humidity, followed by a more pronounced reduction above approximately 60% RH. This trend indicates that the triboelectric output is influenced by ambient humidity at elevated RH.

$$\text{Response } (\%) = [(V_0 - V_{\text{RH}})/V_0] \times 100 \quad (3)$$

To quantify the humidity sensitivity in the high humidity region, the response was calculated according to Equation (3), where $V_0$ is the reference output voltage measured at 51.4% RH, corresponding to the lowest RH included in the selected fitting range under room temperature, and $V_{RH}$ is the output voltage measured at a given RH. As shown in Figure S13b, the response increases monotonically with RH and is well fitted by a second order polynomial ($R^2 = 0.997$). The nonlinear increase in response indicates that the influence of humidity becomes more significant as RH increases. Such behaviour is consistent with the commonly reported effect of adsorbed moisture on triboelectric interfaces, although dedicated studies, including response and recovery time, hysteresis, repeatability, temperature interference, and control sample measurements, are required to measure its efficacy for quantitative humidity sensing.

## Theoretical insights from *ab initio* calculations

Understanding how the dense, nominally nonporous ZIF-zni contributes to triboelectric composites requires examination of its electronic structure and electromechanical response at the atomic scale. Therefore, first-principles density functional theory (DFT) calculations employing the PBE0 hybrid functional were performed, to gain theoretical insights into the triboelectrification behaviour of ZIF-zni.

From DFT we established that, ZIF-zni exhibits a theoretical static dielectric constant value of $\varepsilon' \approx 2.09$, which is in agreement with experimental measurements of a ZIF-zni pellet (2.16 ± 0.07 at 1 MHz, see Figure 3c). This value lies near the upper bound of the typical range reported for ZIFs ($\varepsilon' = 1.6$–$2.2$)[20], however the underlying screening mechanism of ZIF structures is not well understood. Within the idealized model of the periodic unit cell of ZIF-zni, its dense framework ($SAV \approx 1.5\%$) restricts dipole reorientation and exhibits weak frequency-dependent dielectric screening (due to the absence of porosity). These calculations capture intrinsic dielectric screening characteristics that may influence local interfacial electric fields but do not describe the actual ZIF-zni@PU and PDMS interface or quantify interfacial charge transfer. Direct comparisons with porous MOF analogues are required to determine how internal voids and framework flexibility alter this screening behaviour. Frequency-dependent dielectric calculations further identify the THz vibrational modes that are responsible for phonon-assisted polarization (Figure 6a). Imidazolate ring bending (~667 $cm^{-1}$) and Zn-N stretching (~350 $cm^{-1}$) modes exhibit the strongest oscillator strengths and couple efficiently to external strain. In the dense ZIF-zni lattice, these modes respond coherently to mechanical perturbation, which is consistent with possible dynamic polarization during contact events (Movies S3-7). Notably, high-frequency C–H modes contribute negligibly, indicating that the calculated low-frequency dielectric response is dominated by the collective dynamics of the Zn−Im−Zn linkages rather than by flexible organic moieties.

The computed electronic band structure shows a direct bandgap of 7.20 eV (Figure S14a-b), thereby indicating a wide-gap insulating character. This result is consistent with the requirement for an insulating triboelectric layer to hinder charge leakage, although the calculated bandgap should not be interpreted as a measured optical gap. The calculation describes the idealized zni topology without guests, but it does not establish how defects, surfaces, polymer contacts, or adsorbed water modify the experimental interface. The calculated band-edge orbital contributions (Figure S14c) are consistent with prior electronic-structure studies of ZIFs[26,27]. Analysis of the Kohn–Sham wavefunctions (Figure 6b) shows that the valence band maximum is composed primarily of the delocalized $\pi$ and $\sigma$ states spanning the Zn−Im−Zn network, while the conduction band minimum is dominated by the localized antibonding $\sigma^*$ states centered on the tetrahedral nodes comprising $ZnN_4$. These projected density of states (PDOS) and HOCO/LUCO visualizations indicate calculated orbital-localization trends, rather than direct evidence of interfacial charge transfer. Such localization may influence the spatial distribution of electronic polarization in the dense framework.

Furthermore, density functional perturbation theory (DFPT) reveals large calculated Born effective charges ($Z_{Zn} \approx +2.1$ and $Z_N \approx -0.85$), reflecting strong Zn-N covalency and efficient coupling between atomic displacement and polarization. These results suggest that even modest contact-induced strains, particularly at rigid and surface localized coordination motifs, can produce a calculated unit-cell polarization response. The compact, nominally nonporous zni topology may concentrate mechanical deformation into polar Zn-N bonds, although direct comparison with porous analogues is required to quantify this effect. Real-space electrostatic potential maps (Figure 6c) show the predicted electrostatic-potential distribution, revealing sharp alternating potential extrema at Zn coordination sites and imidazolate nitrogen atoms. These calculated local features may be relevant to the surface-accessible ZIF-zni-rich domains observed by nano-FTIR. However, the periodic DFT model does not directly prove or disprove charge transfer at the real ZIF-zni/PU/PDMS interface, which is far more complex than what was modelled above.

Taken together, our theoretical calculations suggest that the high-loading ZIF-zni@PU system may benefit from dense-framework electronic and electromechanical features, including wide-gap insulating behaviour, calculated band-edge asymmetry, large Born effective charges and weak dielectric screening. These periodic calculations offer new insights into potential mechanisms of ZIF dielectrics.

**Discussion**

Beyond the device-level performance, the present study provides broader insight into the design of MOF-enabled triboelectric composites. While triboelectric enhancement has often been associated with porous frameworks possessing large internal surface areas, the dense ZIF-zni framework investigated here has a very low reported solvent-accessible volume and delivers substantial performance enhancement at high filler loading. Nanoscale and multimodal experimental and computational results indicate that triboelectric performance is governed by multiple material descriptors, including dielectric response, surface morphology, framework associated polarization, and work of adhesion, which exhibits an inverse correlation with the triboelectric output. Although the specific contribution of framework porosity cannot be isolated in the present study, the substantial enhancement achieved using the dense ZIF-zni framework is promising. We demonstrate that factors beyond porosity should also be factored into the design of MOF enabled triboelectric materials.

The 20 wt% ZIF-zni@PU composite maintained approximately stable output tested up to ~94,000 cycles under a 100 N cyclic loading, followed by gradual attenuation during extended cycling to 187,000 cycles. The device remained operable during short-term tests under forces approaching 610 N, demonstrating the feasibility of ZIF-zni-based triboelectric composites under realistic mechanical loading conditions. Although the harvested energy remains modest, the device establishes a proof-of-concept platform for distributed energy harvesting and self-powered sensing in smart-flooring environments. The humidity-dependent response is consistent with the interfacial nature of charge generation, where adsorbed water molecules promote charge dissipation and partially screen electrostatic interactions. Notably, this sensitivity is observed despite its very low reported solvent-accessible volume, suggesting that environmental responsiveness can arise from modulation of interfacial contact-electrification processes rather than pore-mediated adsorption.

More broadly, the present work suggests that MOF selection for triboelectric composites may benefit from descriptors beyond porosity, including dielectric response, Born effective charge, band-edge characteristics, and interfacial polarization behaviour. Future comparative studies involving dense and porous frameworks with similar chemistry will be valuable for establishing quantitative structure-property relationships and clarifying the relative contributions of framework density and dielectric screening.

From a practical perspective, the scalable aqueous synthesis of ZIF-zni, together with its high thermal and chemical stability, facilitates scalable filler preparation for triboelectric materials. The integration of energy harvesting and environmental sensing within a single platform highlights the potential of dense MOF composites for multifunctional autonomous systems. More broadly, the present results will motivate further testing of framework-associated interfacial electromechanical effects as a possible design hypothesis for MOF-enabled energy harvesting and self-powered sensing technologies.

**Conclusions and outlook**

In summary, we demonstrate that dense, nominally nonporous ZIF-zni can serve as an effective functional filler for engineering resilient triboelectric composites when being incorporated into a tribopositive PU matrix. The resulting composite exhibits humidity-responsive behaviour and practical use as a triboelectric floor-tile for energy harvesting of pedestrian footsteps.

Combined experimental and theoretical studies indicate that the enhanced triboelectric response is associated with interfacial electro-mechanical effects and modification of surface morphology and local properties.

Notably, our findings revealed that internal surface area of microporous MOF structures should not be treated as the only descriptor of MOF-enabled triboelectric enhancement. Specifically, we have identified dielectric response, polarization susceptibility, and charge-partitioning behaviour as important materials descriptors for triboelectric design. Further investigations of the nanoelectrical factors, in conjunction with fine scale variation of mechanical properties and local chemical response of triboelectric composites are warranted. Beyond the specific ZIF-zni system investigated here, the concept of dense-framework-associated interfacial electromechanical coupling provides another strategy for tailoring triboelectric materials encompassing MOFs as a multifunctional filler.

## Methods

### Materials

All chemicals used in this study are commercially available. Imidazole, zinc nitrate hexahydrate ($Zn(NO_3)_2{\cdot}6H_2O$) and *N*,*N*-dimethylformamide (DMF) were purchased from Fisher Scientific. Polydimethylsiloxane (PDMS) elastomer and curing agent were obtained from Dow Corning (Sylgard 184). A commercial methylene diphenyl isocyanate (MDI)-based polyester/polyether polyurethane (PU) was purchased from Sigma-Aldrich.

### Synthesis of ZIF-zni crystals

Imidazole (3.40 g, 0.05 mol) and zinc nitrate hexahydrate, $Zn(NO_3)_2{\cdot}6H_2O$ (3.69 g, 0.0124 mol), were dissolved in 40 mL and 4 mL of deionised water, respectively. The two solutions were then combined and stirred at room temperature for 45 minutes[28]. The resulting ZIF-zni product was collected by centrifugation and washed three times with deionised water. After air-drying overnight, the sample was activated under vacuum at 100 °C for 3 hours.

### Fabrication of ZIF-zni@PU polymer composites

A 13.7 wt% PU solution was prepared by dissolving 13.7 g of PU in 86.3 g of DMF, followed by stirring at 50 °C overnight. The ZIF-zni mass fraction was defined as wt% = $m_{\text{ZIF-zni}}/(m_{\text{ZIF-zni}}$

+ $m_{PU}$) × 100%. ZIF-zni filler loadings of 2, 5, 10, 15, 20 and 25 wt% were selected to prepare the ZIF-zni@PU composites (i.e., mixed-matrix membranes of MOF@polymer). The ZIF-zni MOF crystals were combined with the 13.7 wt% PU solution using a homogenizer for 10 min before degassing in the vacuum desiccator. The degassed mixtures were then cast onto a TENG substrate, comprising indium tin oxide (ITO)-coated polyethylene terephthalate (PET) sheets, using a doctor-blade with a fixed gap distance (0.8 mm) to ensure a uniform composite film thickness. The cast films were finally dried at 60 °C for 5 h. The nominal thickness of the composite film was found to be below ~100 µm (Table S1).

**Preparation of PDMS membranes**

The PDMS solution was prepared by mixing the precursor and curing agent at a mass ratio of 10:1. The resulting mixture was cast on a ITO-coated PET sheet using a doctor blade with a gap height of 0.8 mm, followed by curing at 90 °C for 3 h.

**Fabrication and measurement of ZIF-zni@PU TENG**

The ZIF-zni@PU-based TENG device was tested under the contact-separation mode. Polymer composites with varying ZIF-zni loadings were cut into 3 cm × 3 cm squares and employed as tribopositive layers, paired against PDMS as the tribonegative layer.

A periodic contact-separation motion was applied using a permanent electromagnetic shaker system (Brüel & Kjær LDS V201) for standard measurement under low load conditions ~100 N, powered by a voltage-amplified function generator (GW Instek AFG-2105) (Figure S1b). In the standardized test, an optimized displacement of 2.5 mm (Figure S3b), a frequency of 2 Hz, and a maximum impact force of ~100 N were used. For large contact force measurements, a larger permanent electromagnetic shaker system (Brüel and Kjær LDS V400) was employed, enabling higher impact forces up to ~610 N (Figure S1c). Measurements were conducted at a frequency of 2 Hz with a fixed separation distance of 2 mm. For comparison between the PU TENG and the 20 wt% ZIF-zni@PU TENG under forces up to ~230 N, a separation distance of 3 mm was used. A load cell (RS PRO 616) integrated into the sample holder was used to monitor the contact force in real time. The instantaneous contact force was maintained at a consistent level by adjusting the separation gap and shaker driving voltage. The open-circuit voltage was recorded using a digital oscilloscope (PicoScope 5444B) with a 100 MΩ high-voltage probe (Rigol RP1300H), while the short-circuit current was measured using a precision electrometer (Keithley 6514). For load-resistance measurements, the voltage across each external resistor was measured using the same TENG device under identical operating conditions.

**Materials characterization**

The surface morphology of the ZIF-zni crystal was revealed by field-emission scanning electron microscope (FESEM) employing TESCAN LYRA3, and the general morphologies of the composite films were analysed through optical microscopy (Olympus Stereo) and digital microscope (Keyence VHX-7000). The crystal structures of the composites were characterised through X-ray diffraction (XRD) using a Rigaku MiniFlex diffractometer with a Cu K$\alpha$ source (1.541 Å). The Fourier-transform infrared (FTIR) spectra were recorded by a Nicolet iS10 FTIR spectrometer equipped with an attenuated total reflectance (ATR) module.

Synchrotron-based far-infrared (FIR) measurements employing an ATR module were conducted at beamline B22 MIRIAM of Diamond Light Source (DLS). Measurements were performed on the Bruker VERTEX 80V FTIR bench equipped with a liquid helium-cooled bolometer detector. ZIF-zni powder was loaded in small quantities (~mg) using a spatula onto the ATR crystal for THz measurements in ATR-FTIR mode.

Micro-Raman spectroscopy was performed using a Renishaw inVia Raman microscope equipped with 532 nm and 785 nm excitation lasers. The low-frequency Raman region (5-400 $cm^{-1}$) was measured using a 532 nm excitation laser, an 1800 grooves $mm^{-1}$ grating, and an Eclipse notch filter with a 5 $cm^{-1}$ cutoff. The higher frequency region (600-1510 $cm^{-1}$) was measured using a 785 nm excitation laser, a 1200 grooves $mm^{-1}$ grating, and an edge filter. Data were collected using a 50× objective lens at 10% laser power. For Raman mapping, a 20 µm × 20 µm area was scanned, with markers used to ensure consistent positioning for both the ZIF-zni@PU matrix and PDMS samples before and after extended contact-separation tests.

Atomic force microscopic (AFM) surface height topography and nano-FTIR spectra *via* nearfield infrared nanospectroscopy were determined through a scattering-type scanning near-field optical microscope (Neaspec s-SNOM)[29]. The nano-FTIR spectra were taken at ~20 nm spatial resolution, using a Toptica differential frequency generation (DFG) laser tuned to a frequency range of 1100 to 1900 $cm^{-1}$. Near-field pseudoheterodyne (PsHet) nanoimaging was performed using a widely-tunable optical parametric oscillator laser (Piano wOPO, Stuttgart

Instruments) as the infrared excitation source, coupled to the s-SNOM operated in tapping mode. An electrically conductive platinum iridium-coated AFM probe (Arrow-NCPt) with a nominal tip radius below 25 nm and a resonance frequency of 285 kHz was employed. Background calibration was carried out using a TGQ-1 silicon/$SiO_2$ standard. Imaging was conducted at 1496 $cm^{-1}$, corresponding to the vibrational mode of the imidazolate linker in ZIF-zni. Both the amplitude (A) and phase (P) signals were collected at higher-order optical harmonics, each probing different near-field interaction depths. The probing depth is approximately 40 nm for the second harmonic O2, and 20 nm for the third harmonic O3. Normalization of the optical signals determined from the different harmonics, such as O3A/O2A and O3P-O2P, was applied to suppress far-field background contributions, including scattering, reflections, and geometric artefacts[30].

The Kelvin-probe force microscopy (KPFM) data were recorded using an Asylum Research Cypher ES equipped with an ASYELEC-01-R2 conductive tip under SKPM mode. The scan area was set to 5 μm × 5 μm, with a scan resolution of 128 × 128 pixels and a scan rate of 2 Hz to minimize drift. The KPFM results were processed using the Gwyddion software[31], where the colour scale and range were adjusted, and the average surface potential was calculated. No reference-standard calibration was performed; therefore, the extracted work-function values are reported as relative estimates to that of the neat PU.

Pull-off test was conducted in an instrumented nanoindenter (iMicro KLA-Tencor), employing a cylindrical flat punch with a nominal diameter of 10 μm. Initially, the flat punch was positioned 2 μm above the sample surface[25]. Then, the tip approached the surface at a speed of 100 nm $s^{-1}$, with surface detection facilitated by monitoring the phase signal. After contact, the punch took 10 s to load until reaching a maximum load of 0.1 mN. The tip was held at this peak load for 2 s. Then the tip was unloaded at the same rate and retracted 5 μm from the surface. The pull-off stress was calculated by dividing the measured pull-off force by the nominal contact area of the flat punch. The work of adhesion was calculated by integrating the unloading/retraction segment of the load–depth curve and normalizing the result by the nominal punch area; data were reported as mean ± standard deviation (SD) from three measurements.

Dielectric measurements were assessed by an inductance-capacitance-resistance (LCR) meter (Hioki IM3536) as a function of frequency sweeping from 4 Hz to 8 MHz. A parallel-plate capacitor configuration was used for the dielectric measurements[32]. For ZIF-zni pellet measurements, the two opposing faces of each 1 cm diameter pellet were coated with a thin gold layer using a sputter coater (Quorum SC7620). For the composite films (without ITO-PET), capacitance was measured across the film thickness in a parallel-plate configuration, and the relative permittivity was calculated as $\varepsilon_r = Cd/(\varepsilon_0 A)$, using the measured dry-film thickness $d$ and nominal electrode area $A$. Values at 1 MHz are reported as mean ± SD from three repeat measurements performed on a single pellet (Table S2).

**Acknowledgements**

This work was supported by the EPSRC Frontier Research Grant (TEGMOF grant EP/Z534146/1). We thank Tianhuai Xu, Dr. Mark Frogley and Dr. Alex Hawkins for help on the synchrotron-based FIR measurements at B22 MIRIAM in Diamond Light Source *via* beamtime SM40142. We thank Dr Svemir Rudić for enabling access to the SCRAF high-performance computing cluster at STFC Rutherford Appleton Laboratory in Harwell Campus. D.M. acknowledges the University of Oxford Advanced Research Computing (ARC) facility and the ARCHER2 UK National Supercomputing Service through the UK HEC Materials Chemistry Consortium, which is funded by EPSRC grant EP/R029431/1.

**Author contributions**

C.Z.J. conceived the project with input from J.C.T., prepared the samples, performed all data analysis, and wrote the first draft of the manuscript; D.M. conducted the DFT computational studies; J.Y. helped to develop and test the applications; J.C.T. supervised C.Z.J., revised the draft manuscript, discussed the results and contributed to the final version of the manuscript with C.Z.J.

**Competing interests**

The authors declare no competing interests.

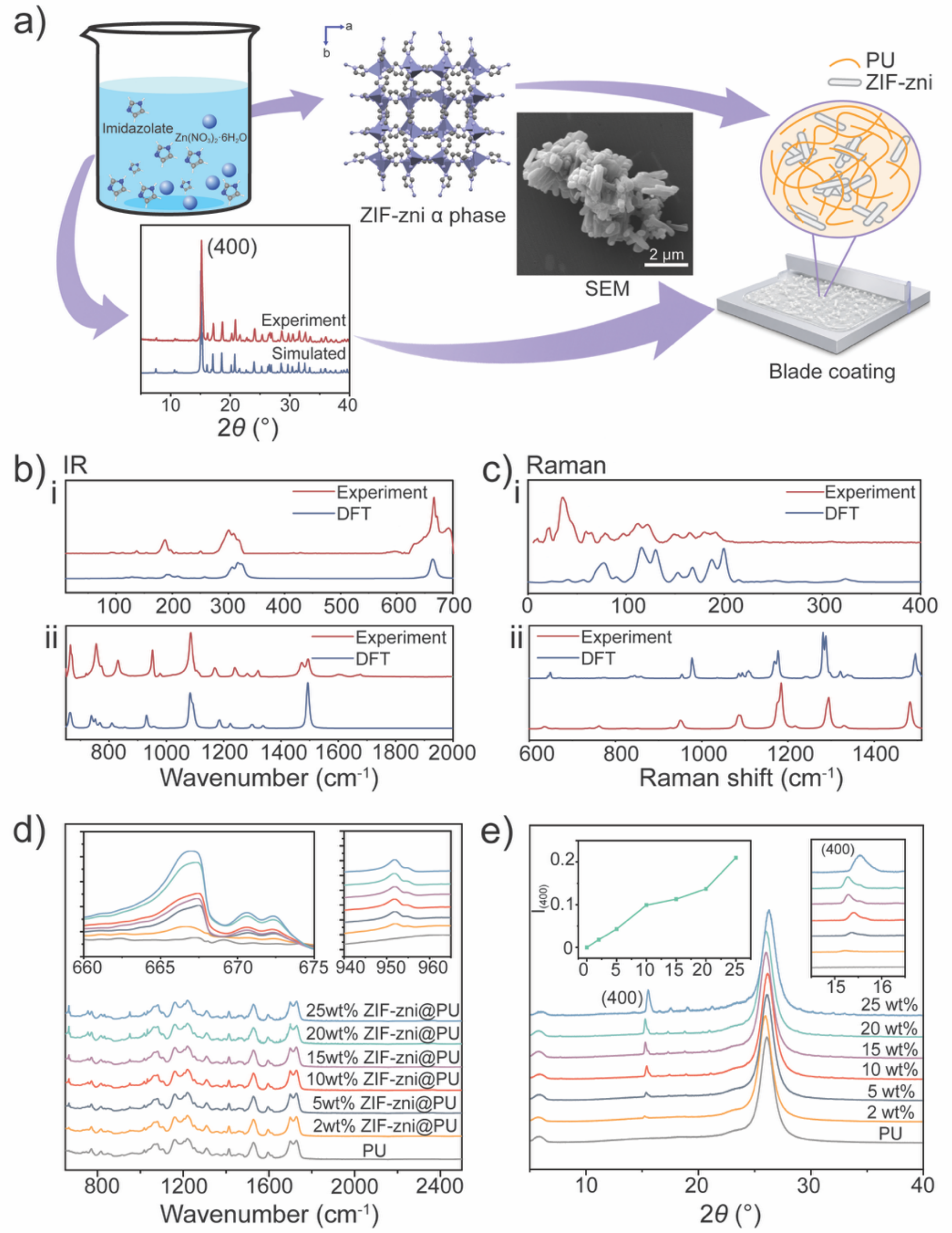

**Figure 1.** Structural characterization of ZIF-zni@PU composites. **a)** Schematic illustration of the fabrication process of ZIF-zni@PU composite films, together with the crystal structure, SEM image, and XRD pattern of the as-synthesized ZIF-zni. **b)** Experimental and DFT-simulated IR spectra of ZIF-zni (DFT scaling factor = 0.95) including i) far-infrared spectra in the 0-700 $cm^{-1}$ range, and ii) ATR-FTIR/mid-infrared spectra in the 650-2000 $cm^{-1}$ range. **c)** Comparison of experimental and DFT-simulated (scaling factor = 0.95) Raman spectra of ZIF-zni in the **i)** ranges 0-400 $cm^{-1}$ and **ii)** 600-1510 $cm^{-1}$. **d)** ATR-FTIR spectra of PU and ZIF-zni@PU composites with different wt.% loadings of ZIF-zni fillers. Insets highlight the characteristic bands at 668 $cm^{-1}$ and 950 $cm^{-1}$. **e)** XRD patterns of ZIF-zni@PU composites with different ZIF-zni wt.% loadings. Insets show an enlarged view of the (400) reflection and the corresponding peak intensity as a function of ZIF-zni wt.% loading.

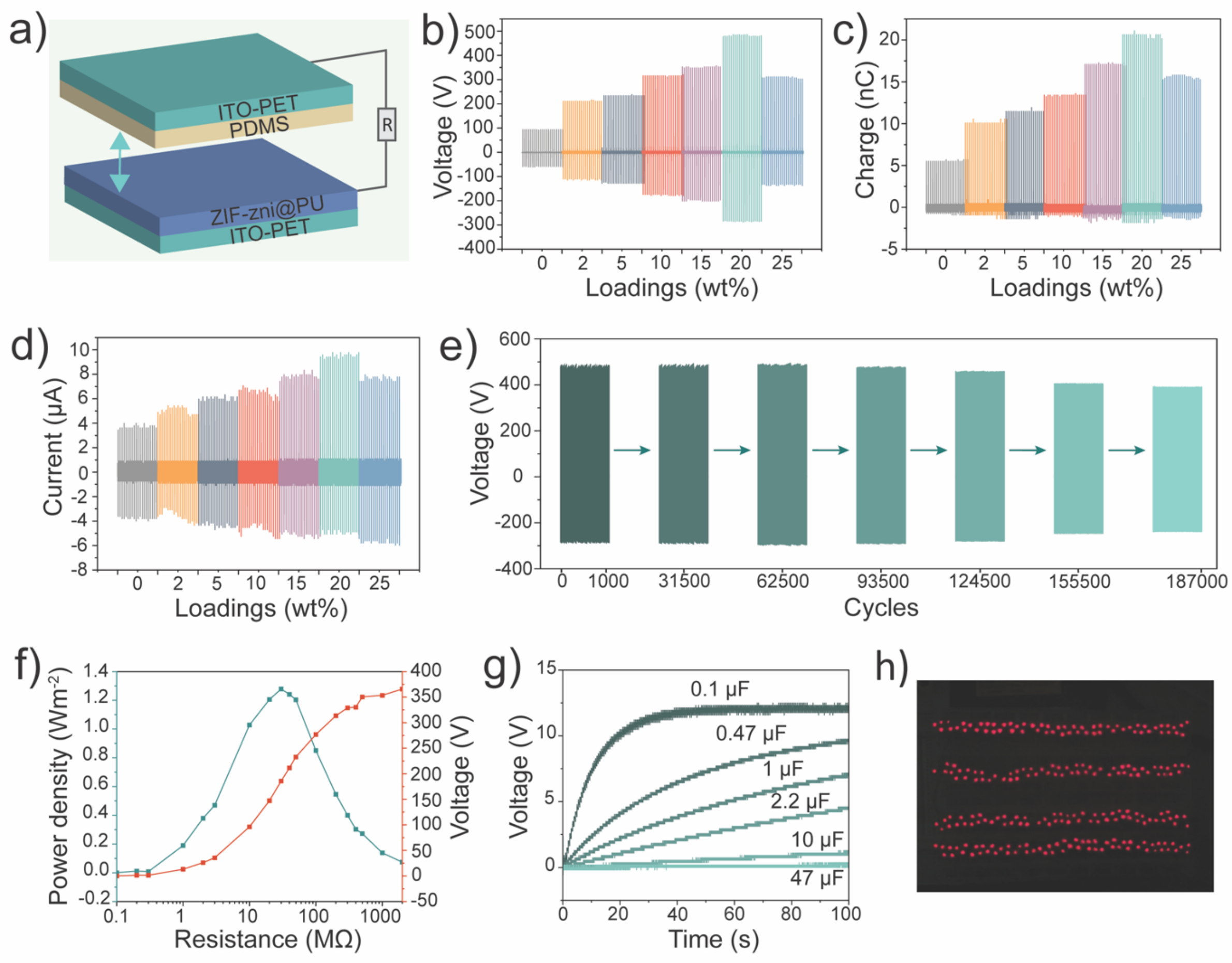

**Figure 2. Electrical output performance of ZIF-zni@PU triboelectric nanogenerators. a)** Schematic illustration of the ZIF-zni@PU against PDMS TENG device operating in contact–separation mode. **b-d)** Open-circuit voltage ($V_{oc}$), transferred charge ($Q_{sc}$), and short-circuit current ($I_{sc}$) of ZIF-zni@PU TENGs with different ZIF-zni wt% loadings. Each ZIF-zni loading is represented by a 10 s interval. **e)** Durability of the 20 wt% ZIF-zni@PU TENG over 187,000 operation cycles. **f)** Peak output power density and corresponding load voltage as a function of external load resistance for the 20 wt% ZIF-zni@PU TENG. **g)** Charging profiles of capacitors with different capacitances powered by the 20 wt% ZIF-zni@PU TENG operating at 2 Hz. **h)** Photograph of 240 illuminated LEDs powered by the 20 wt% ZIF-zni@PU TENG operated at 2 Hz with an applied force of approximately 100 N and a separation distance of 2.5 mm.

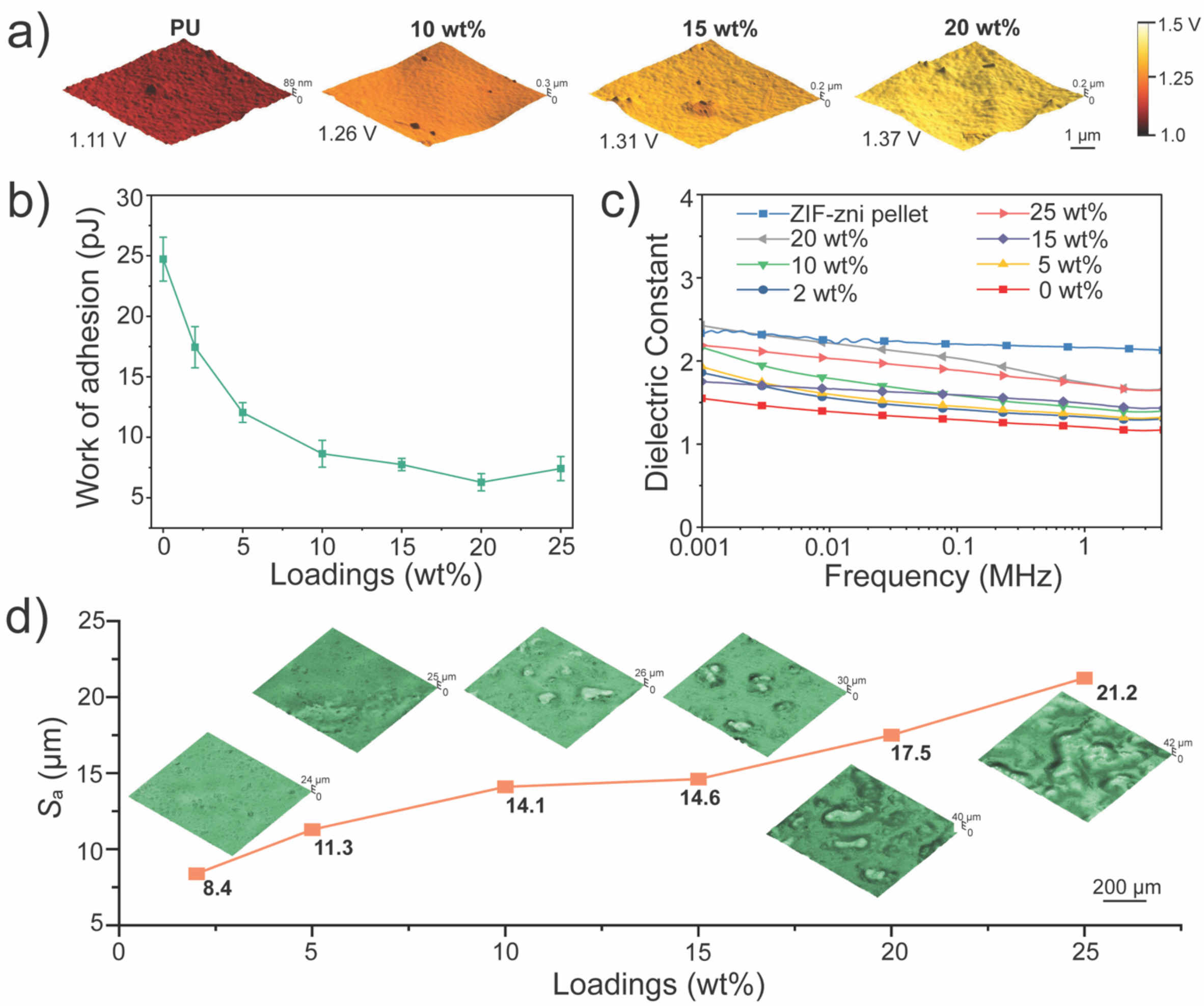

**Figure 3. Surface properties of ZIF-zni@PU composite films with different ZIF-zni filler loadings. a)** Representative KPFM surface potential maps (5 μm × 5 μm) with average values of surface potentials obtained from pristine PU and composite films, containing 10, 15 and 20 wt% of ZIF-zni; **b)** work of adhesion with different ZIF-zni loadings; **c)** Dielectric constants of ZIF-zni pellet and varied loading of ZIF-zni@PU films; d) Representative 3D surface topography images and corresponding root mean square roughness ($S_q$) values measured over an area of 740 μm × 500 μm.

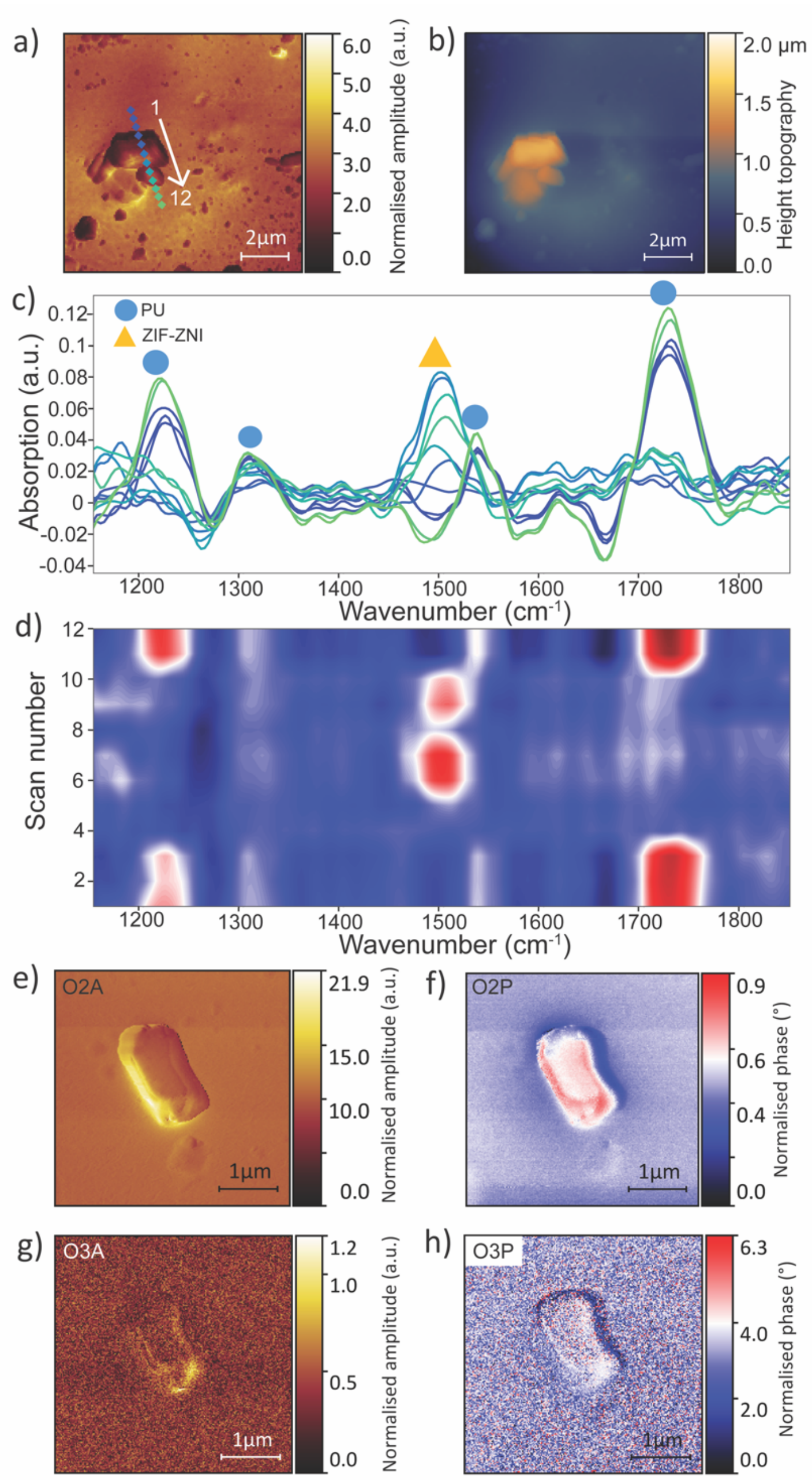

**Figure 4. Nanoscale characterization of ZIF-zni crystals in ZIF-zni@PU composites. a)** Nano-FTIR optical amplitude image (O2A) showing the line scan path and 12 sampling positions. **b)** Height topography image of a surface-exposed ZIF-zni crystal. **c)** Nano-FTIR spectra acquired from the 12 positions indicated in panel a. **d)** Nano-FTIR spectral contour map constructed from the line scans shown in panel a. **e–h)** Near-field PsHet images at 1496 $cm^{-1}$ for a representative ZIF-zni crystal deposited on a flat silicon substrate, including second-harmonic amplitude (O2A), second-harmonic phase (O2P), third-harmonic amplitude (O3A), and third-harmonic phase (O3P) channels.

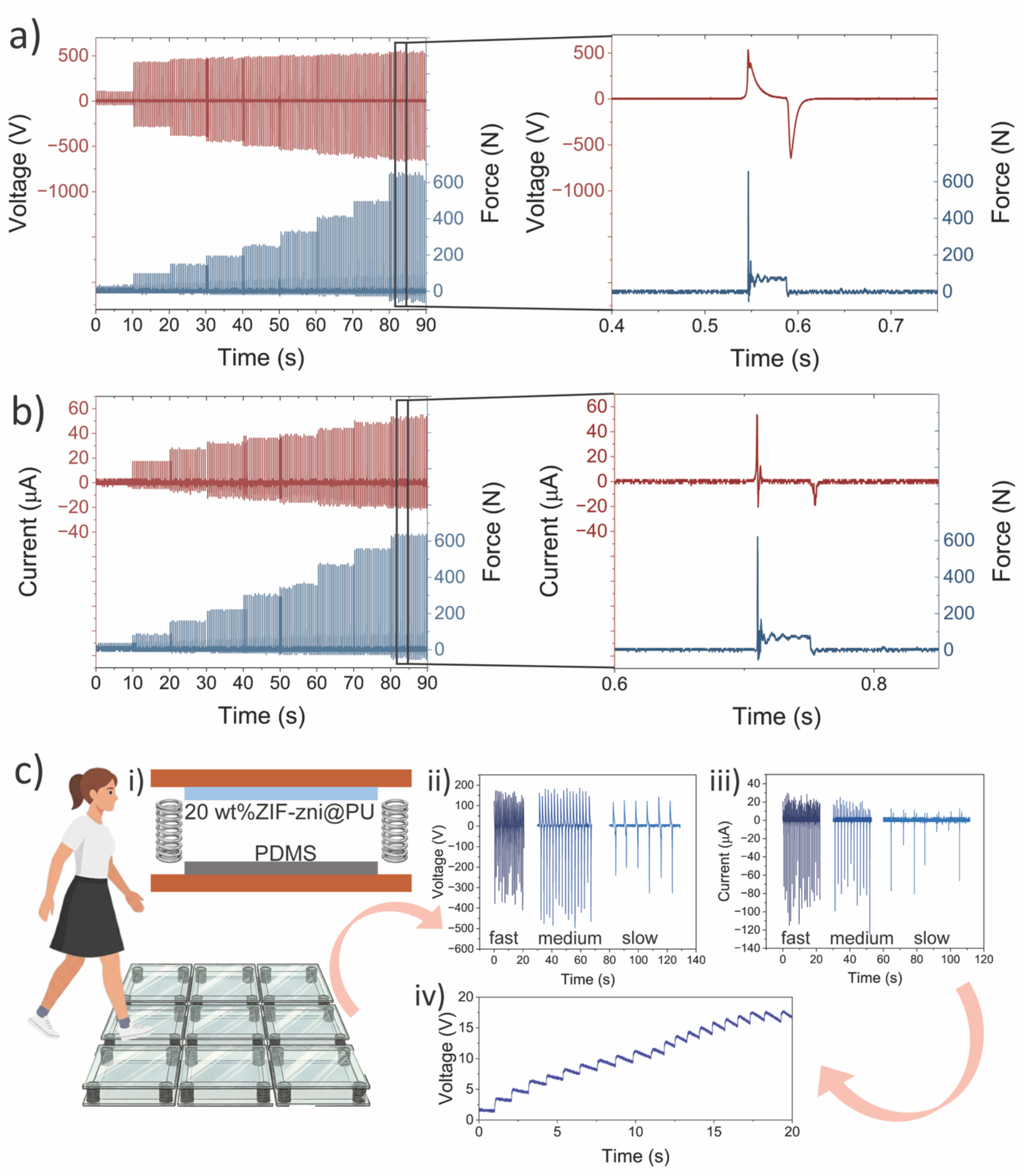

**Figure 5. Application of the 20 wt% ZIF-zni@PU composite film as a compact floor tile energy harvester. a)** Voltage output under a maximum applied force of 610 N with enlarged view of a representative voltage peak from **a)**. **b)** The corresponding current output measured from the same film with enlarged view of a representative current peak from **b)**, **c)** Demonstration of **i)** a 20 cm × 20 cm floor tile module incorporating the composite film, showing human foot-step response (~60 kg normal load), **ii)** voltage and **iii)** current output at fast (~1 step $s^{-1}$), medium (~0.5 step $s^{-1}$), and slow (~0.12 step $s^{-1}$) stepping rates, and **iv)** the charging of a 1 μF capacitor under a medium pace.

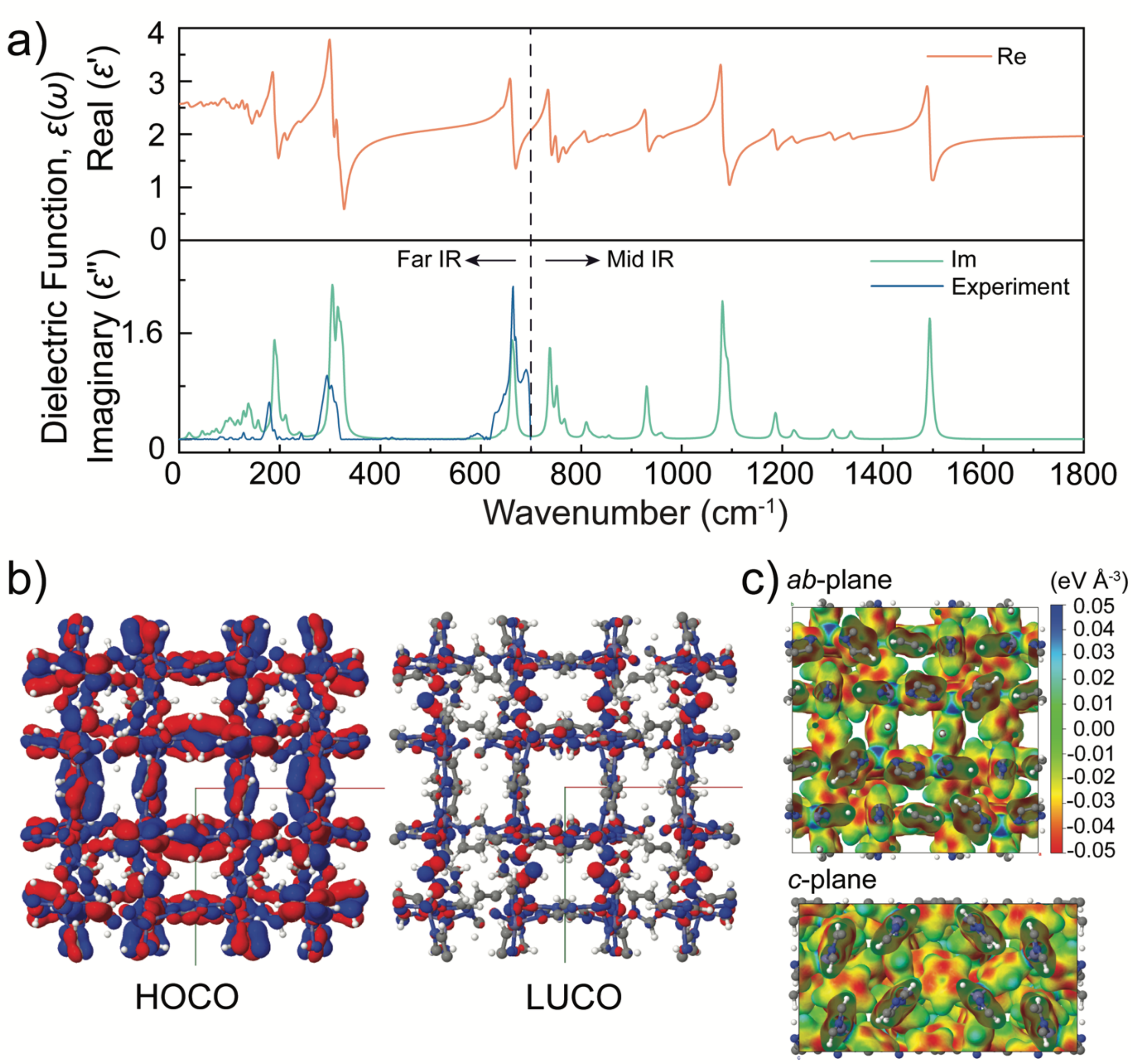

**Figure 6. Electronic structure and dielectric response of ZIF-zni from DFT calculations.** a) Computed frequency-dependent dielectric function $\varepsilon(\omega)$, where the real part $\mathrm{Re}[\varepsilon(\omega)] = \varepsilon'$ and the imaginary part $\mathrm{Im}[\varepsilon(\omega)]$ = absorption $\varepsilon''$ (shift factor: 0.95), together with experimental FIR data. Distinct features originate from imidazolate bending and Zn-N stretching vibrations, which govern the low-frequency polarization response of the dense framework. b) Calculated isosurfaces of the highest occupied and lowest unoccupied crystalline orbitals (HOCO and LUCO), showing calculated linker-dominated states at the VB edge and localized metal–ligand states at the CB edge; these are orbital-localization trends rather than direct evidence of interfacial charge transfer. c) Electrostatic potential maps projected along the crystallographic *ab*-plane and *c*-plane, illustrating alternating potential maxima at the Zn coordination sites (blue) and minima around the imidazolate nitrogen atoms (red).

## References

1 Wu, C., Wang, A. C., Ding, W., Guo, H. & Wang, Z. L. Triboelectric Nanogenerator: A Foundation of the Energy for the New Era. *Adv. Energy Mater.* **9**, 1802906 (2019). https://doi.org:10.1002/aenm.201802906

2 Zhang, C. *et al.* Advances in High-Performance Energy Harvesting TENG and its Applications. *Adv. Funct. Mater.* **36**, e28299 (2026). https://doi.org:10.1002/adfm.202528299

3 Li, Y. *et al.* Advanced Dielectric Materials for Triboelectric Nanogenerators: Principles, Methods, and Applications. *Adv. Mater.* **36**, 2314380 (2024). https://doi.org:10.1002/adma.202314380

4 Chen, A., Zhang, C., Zhu, G. & Wang, Z. L. Polymer Materials for High-Performance Triboelectric Nanogenerators. *Adv. Sci.* **7**, 2000186 (2020). https://doi.org:10.1002/advs.202000186

5 Du, T. *et al.* Advances in Green Triboelectric Nanogenerators. *Adv. Funct. Mater.* **34** (2024). https://doi.org:10.1002/adfm.202313794

6 Choi, Y. S., Kim, S. W. & Kar-Narayan, S. Materials-Related Strategies for Highly Efficient Triboelectric Energy Generators. *Adv. Energy Mater.* **11**, 2003802 (2021). https://doi.org:10.1002/aenm.202003802

7 Cao, X., Jie, Y., Wang, N. & Wang, Z. L. Triboelectric Nanogenerators Driven Self-Powered Electrochemical Processes for Energy and Environmental Science. *Adv. Energy Mater.* **6**, 1600665 (2016). https://doi.org:10.1002/aenm.201600665

8 Yi, F., Zhang, Z., Kang, Z., Liao, Q. & Zhang, Y. Recent Advances in Triboelectric Nanogenerator-Based Health Monitoring. *Adv. Funct. Mater.* **29**, 1808849 (2019). https://doi.org:10.1002/adfm.201808849

9 Wang, Y.-M. *et al.* Advances in Metal–Organic Framework-Based Triboelectric Nanogenerators. *ACS Mater. Lett.* **6**, 3883-3898 (2024). https://doi.org:10.1021/acsmaterialslett.4c00773

10 Jin, C. & Tan, J.-C. Robust triboelectric energy harvesters engineered from electrochemically deposited films of HKUST-1 polycrystals. *Commun. Chem.* **9** (2026). https://doi.org:10.1038/s42004-026-01949-0

11 Wang, Y. *et al.* A Tribo/Piezoelectric Nanogenerator Based on Bio-MOFs for Energy Harvesting and Antibacterial Wearable Device. *Adv. Mater.* **37** (2025). https://doi.org:10.1002/adma.202418207

12 Hajra, S. *et al.* A Green Metal–Organic Framework-Cyclodextrin MOF: A Novel Multifunctional Material Based Triboelectric Nanogenerator for Highly Efficient Mechanical Energy Harvesting. *Adv. Funct. Mater.* **31**, 2101829 (2021). https://doi.org:10.1002/adfm.202101829

13 Wang, Y.-M. *et al.* Remarkable improvement of MOF-based triboelectric nanogenerators with strong electron-withdrawing groups. *Nano Energy* **107**, 108149 (2023). https://doi.org:10.1016/j.nanoen.2022.108149

14 Wen, R., Fan, L., Li, Q. & Zhai, J. A composite triboelectric nanogenerator based on flexible and transparent film impregnated with ZIF-8 nanocrystals. *Nanotechnology* **32**, 345401 (2021). https://doi.org:10.1088/1361-6528/ac020f

15 Wang, Y.-M. *et al.* Highly stable metal-organic framework UiO-66-$NH_2$ for high-performance triboelectric nanogenerators. *Nanotechnology* **33**, 065402 (2022). https://doi.org:10.1088/1361-6528/ac32f8

16 Ye, J. & Tan, J.-C. High-performance triboelectric nanogenerators incorporating chlorinated zeolitic imidazolate frameworks with topologically tunable dielectric and surface adhesion properties. *Nano Energy* **114**, 108687 (2023). https://doi.org:10.1016/j.nanoen.2023.108687
17 Pandey, P. *et al.* Metal-organic frameworks-based triboelectric nanogenerator powered visible light communication system for wireless human-machine interactions. *Chem. Eng. J.* **452**, 139209 (2023). https://doi.org:10.1016/j.cej.2022.139209
18 Zhou, J. *et al.* Breathable Metal–Organic Framework Enhanced Humidity-Responsive Nanofiber Actuator with Autonomous Triboelectric Perceptivity. *ACS Nano.* **17**, 17920-17930 (2023). https://doi.org:10.1021/acsnano.3c04022
19 Tan, J. C., Bennett, T. D. & Cheetham, A. K. Chemical structure, network topology, and porosity effects on the mechanical properties of Zeolitic Imidazolate Frameworks. *Proc. Natl. Acad. Sci. U. S. A.* **107**, 9938-9943 (2010). https://doi.org:10.1073/pnas.1003205107
20 Ryder, M. R. *et al.* Dielectric Properties of Zeolitic Imidazolate Frameworks in the Broad-Band Infrared Regime. *J. Phys. Chem. Lett.* **9**, 2678-2684 (2018). https://doi.org:10.1021/acs.jpclett.8b00799
21 Naseri, M. *et al.* Lipase and Laccase Encapsulated on Zeolite Imidazolate Framework: Enzyme Activity and Stability from Voltammetric Measurements. *ChemCatChem* **10**, 5425-5433 (2018). https://doi.org:10.1002/cctc.201801293
22 Wen, R., Guo, J., Yu, A., Zhai, J. & Wang, Z. L. Humidity-Resistive Triboelectric Nanogenerator Fabricated Using Metal Organic Framework Composite. *Adv. Funct. Mater.* **29**, 1807655 (2019). https://doi.org:10.1002/adfm.201807655
23 Das, N. K., Ravipati, M. & Badhulika, S. Nickel Metal-Organic Framework/PVDF Composite Nanofibers-based Self-Powered Wireless Sensor for Pulse Monitoring of Underwater Divers via Triboelectrically Generated Maxwell's Displacement Current. *Adv. Funct. Mater.* **33** (2023). https://doi.org:10.1002/adfm.202303288
24 Guo, Y. *et al.* in *Fluorinated metal-organic framework as bifunctional filler toward highly improving output performance of triboelectric nanogenerators* (Elsevier, 2020).
25 Xu, T., Ye, J. & Tan, J. C. Unravelling the Ageing Effects of PDMS - Based Triboelectric Nanogenerators. *Adv. Mater. Interfaces* **11**, 2400094 (2024). https://doi.org:10.1002/admi.202400094
26 Mandal, A., Khuntia, S. K., Mondal, D., Mahadevan, P. & Bhattacharyya, S. Spin Texture Sensitive Photodetection by Dion–Jacobson Tin Halide Perovskites. *J. Am. Chem. Soc.* **145**, 24990-25002 (2023). https://doi.org:10.1021/jacs.3c10195
27 Mondal, D. & Mahadevan, P. Shape Control of Emissive Properties of Mn-Doped CsPbBr 3 Nanocrystals. *J. Phys. Chem.* **125**, 11462-11467 (2021). https://doi.org:10.1021/acs.jpcc.1c02368
28 Tocco, D. *et al.* Enzyme immobilization on metal organic frameworks: Laccase from Aspergillus sp. is better adapted to ZIF-zni rather than Fe-BTC. *Colloids Surf. B Biointerfaces* **208**, 112147 (2021). https://doi.org:10.1016/j.colsurfb.2021.112147
29 Möslein, A. F., Gutiérrez, M., Cohen, B. & Tan, J.-C. Near-Field Infrared Nanospectroscopy Reveals Guest Confinement in Metal–Organic Framework Single Crystals. *Nano Lett.* **20**, 7446-7454 (2020). https://doi.org:10.1021/acs.nanolett.0c02839
30 Mester, L., Govyadinov, A. A. & Hillenbrand, R. High-fidelity nano-FTIR spectroscopy by on-pixel normalization of signal harmonics. *Nanophotonics* **11**, 377-390 (2022). https://doi.org:10.1515/nanoph-2021-0565

31 Nečas, D. & Klapetek, P. Gwyddion: an open-source software for SPM data analysis. *Open Phys.* **10**, 181-188 (2012). https://doi.org:10.2478/s11534-011-0096-2

32 Babal, A. S. *et al.* Impact of Pressure and Temperature on the Broadband Dielectric Response of the HKUST-1 Metal–Organic Framework. *J. Phys. Chem. C* **123**, 29427-29435 (2019). https://doi.org:10.1021/acs.jpcc.9b08125

## *Supporting Information*

## *for*

# Triboelectrification of a dense metal-organic framework for resilient mechanical energy harvesters

*Chuzhan Jin, Debayan Mondal, Jiahao Ye, Jin-Chong Tan*[*]

Multifunctional Materials & Composites (MMC) Laboratory, Department of Engineering Science, University of Oxford, Parks Road, Oxford OX1 3PJ, U.K.

[*]Corresponding Author:
jin-chong.tan@eng.ox.ac.uk

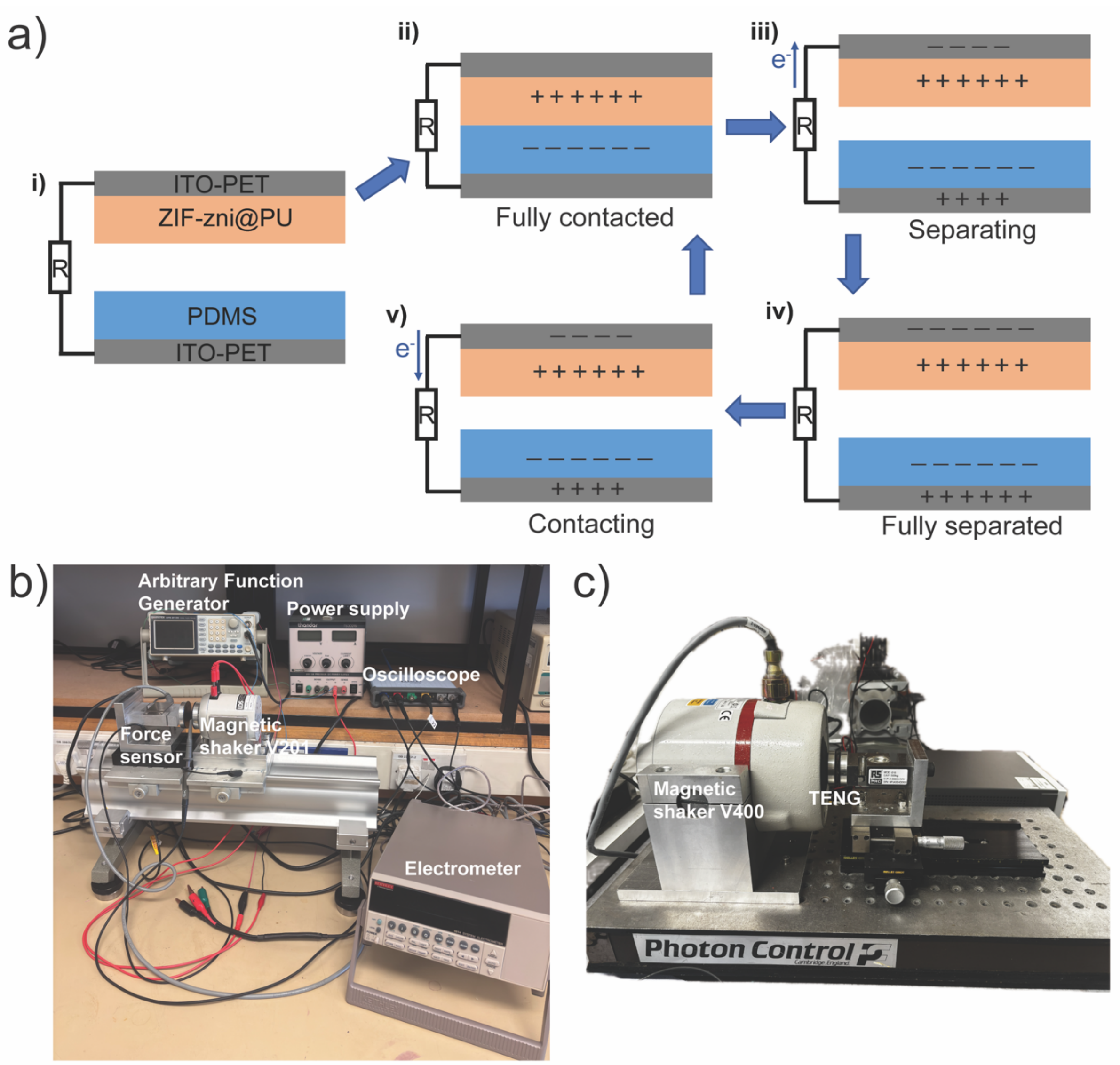

**Figure S1.** a) Schematic diagram of the proposed working mechanism of the ZIF-zni@PU against PDMS TENG device under the contact-separation mode; Photos of TENG set-up under contact-separation mode using b) a smaller electromagnetic shaker V201 for force of ~100 N, and c) a large electromagnetic shaker V400 to yield higher contact force of up to ~ 600 N.

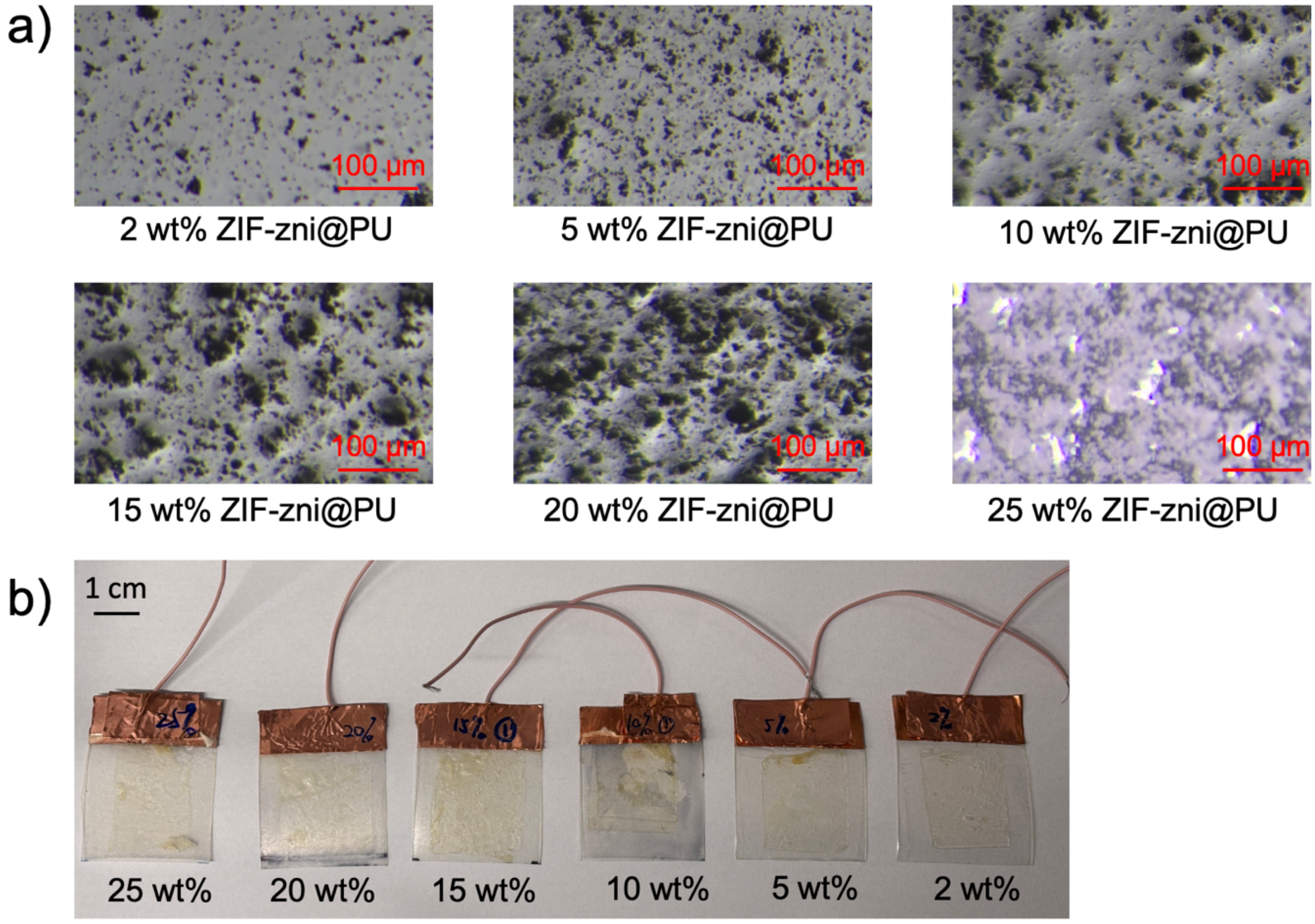

**Figure S2.** Images of ZIF-zni@PU composite samples. a) Optical stereomicroscopy images of PU composites with different ZIF-zni filler loadings in wt%. b) Photographs of devices comprising PU composite films with a systematically varied ZIF-zni loading content from 2–25 wt%.

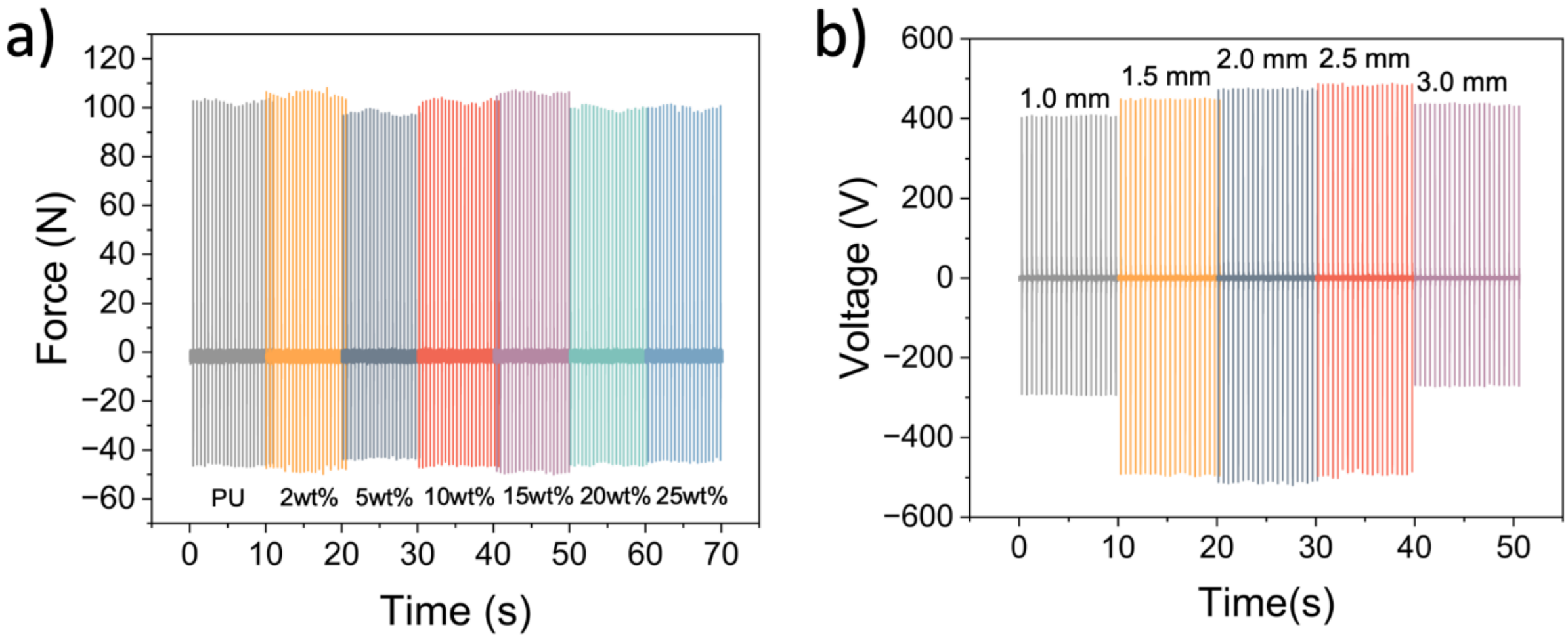

**Figure S3.** a) Force measurements of the ZIF-zni@PU TENG vs PDMS under contact-separation tests. b) Open-circuit voltage determined using various contact-separation distance tested using the 20 wt% ZIF-zni@PU TENG vs PDMS.

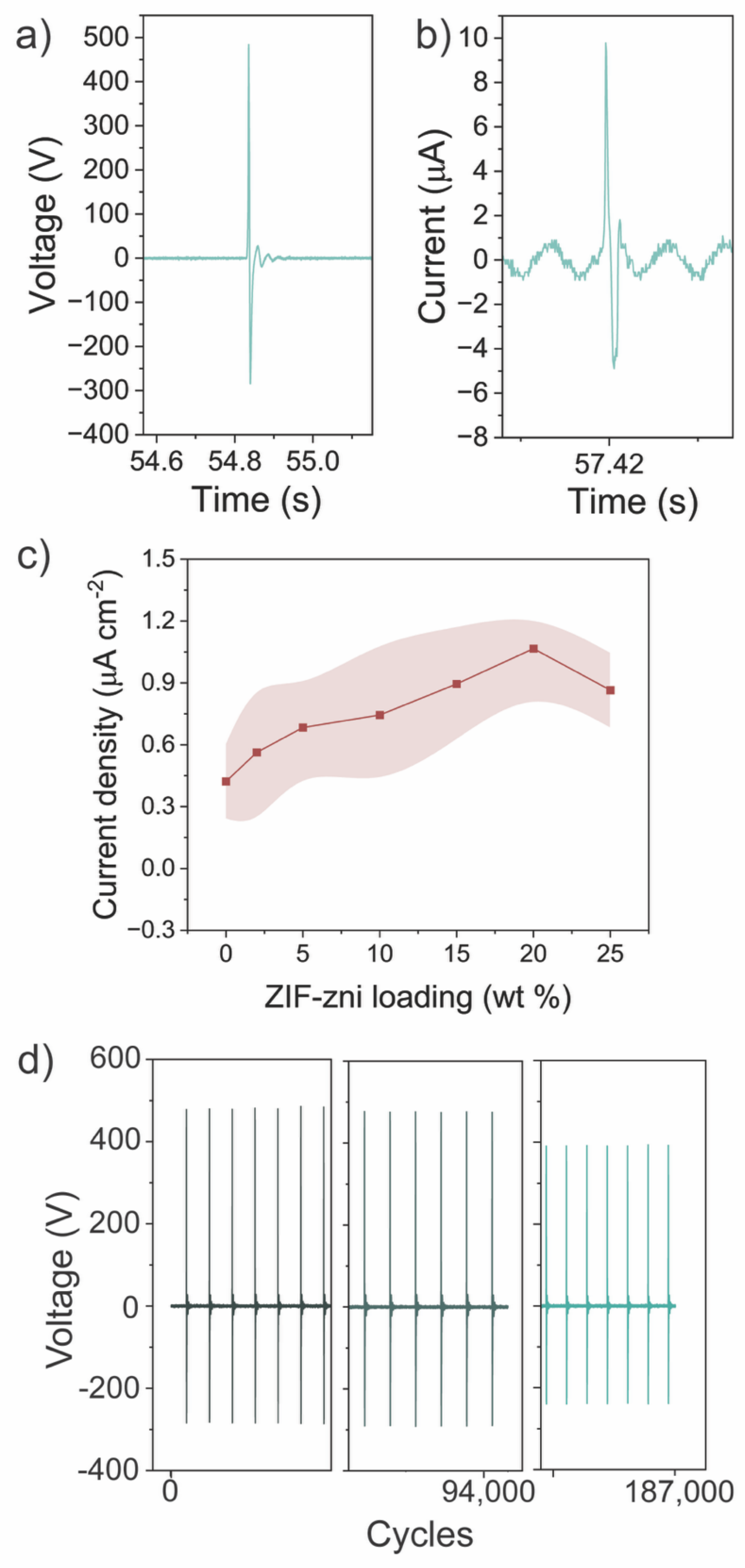

**Figure S4.** ZIF-zni@PU TENG vs PDMS performance. Single cycle of output a) open-circuit voltage and b) short-circuit current of 20 wt% ZIF-zni@PU. c) Current density at different ZIF-zni loadings (wt%). The shaded region in c) represents the minimum-to-maximum range across repeated measurements. d) Magnified view of a long-term stability test starting from 0 cycle, intermediate stage at approximately 94,000 cycles, and at the end of 187,000 cycles.

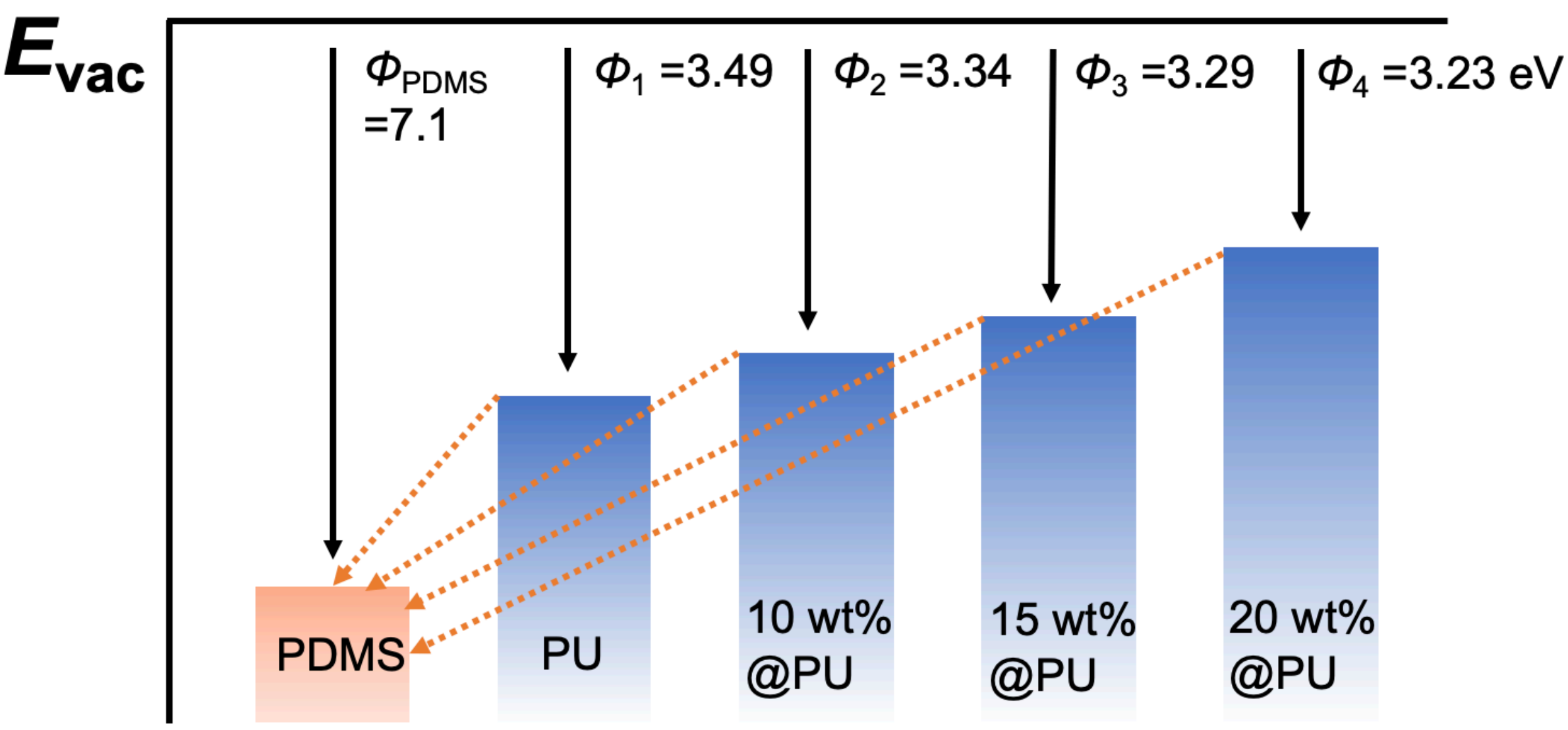

**Figure S5.** Work-function alignment diagram of ZIF-zni@PU TENG, where $E_{vac}$ denotes the vacuum energy level. The estimated work function was calculated from the contact potential difference according to $\Phi_{sample}$ (eV) = $\Phi_{tip}$ (eV) − $V_{CPD}$ (V), using the reported 4.6 eV work function of the Pt/Ir-coated AFM tip. The diagram is intended for qualitative comparison of relative trends rather than calibrated absolute work-function values.

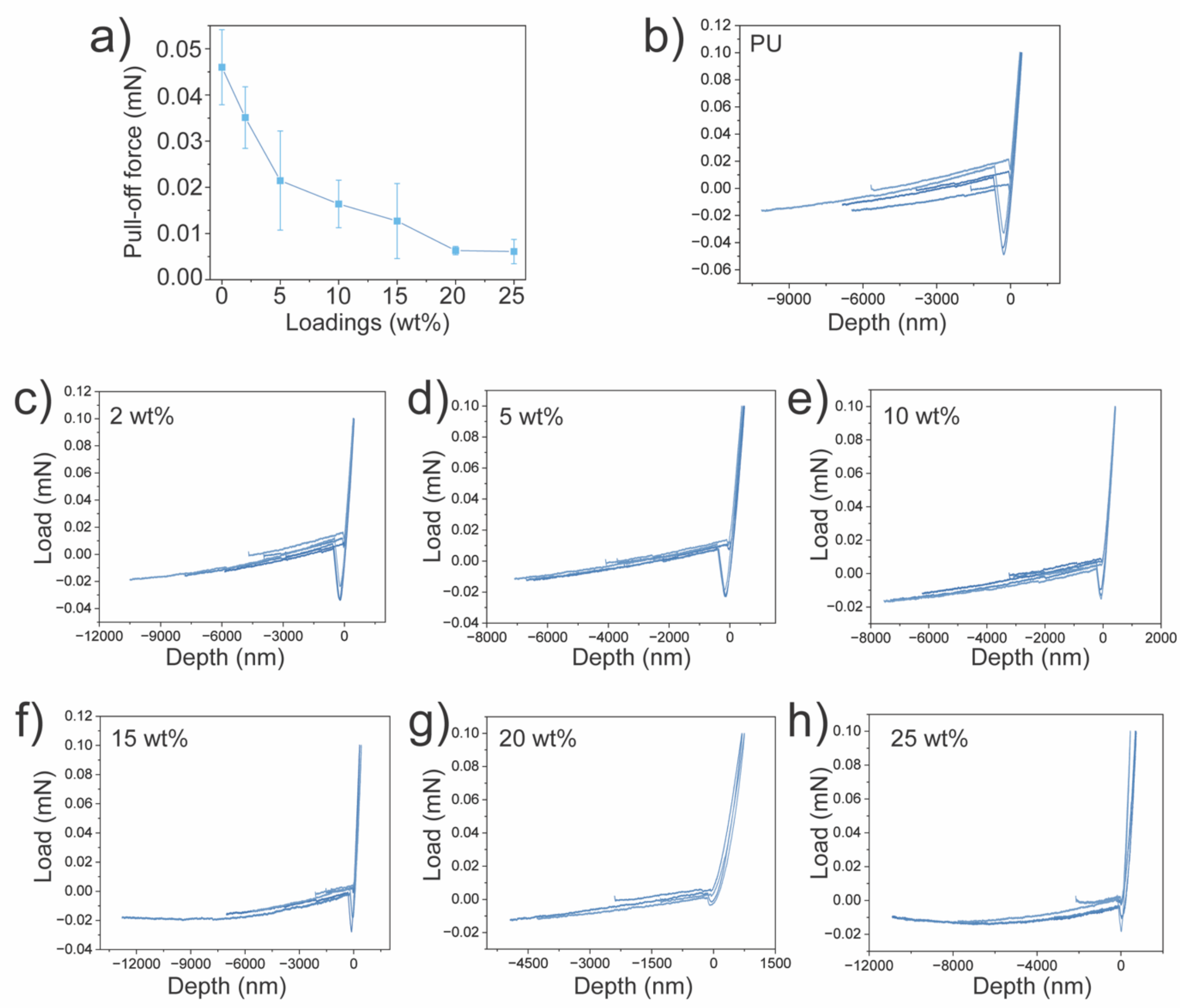

**Figure S6.** a) Pull-off force among different loadings of ZIF-zni@PU composite films. The measurements were performed using a nanoindenter equipped with a 10 µm diameter cylindrical flat punch; data are reported as mean ± SD from 3 measurements. b-h) Representative load–displacement curves of pristine PU and ZIF-zni@PU composite films containing 2, 5, 10, 15, 20, and 25 wt% ZIF-zni, respectively, used to determine the pull off force and work of adhesion.

**Table S1.** Average thickness of ZIF-zni@PU composite films, reported as mean ± SD from 3 measurements.

| ZIF-zni loading (wt%) | Thickness (mm) |
|---|---|
| PU | 0.085 ± 0.002 |
| 2 wt% ZIF-zni@PU | 0.087 ± 0.002 |
| 5 wt% ZIF-zni@PU | 0.093 ± 0.003 |
| 10 wt% ZIF-zni@PU | 0.093 ± 0.002 |
| 15 wt% ZIF-zni@PU | 0.089 ± 0.002 |
| 20 wt% ZIF-zni@PU | 0.092 ± 0.001 |
| 25 wt% ZIF-zni@PU | 0.098 ± 0.004 |

**Table S2.** Dielectric constant of varied loading of ZIF-zni@PU film measured from LCR meter employing a parallel-plate capacitor setup, reported as mean ± SD at 1 MHz from 3 measurements.

| ZIF-zni loading (wt%) | Dielectric constant, $\varepsilon'$ (-) at 1 MHz |
|---|---|
| PU | 1.23 ± 0.02 |
| 2 wt% ZIF-zni@PU | 1.33 ± 0.03 |
| 5 wt% ZIF-zni@PU | 1.38 ± 0.02 |
| 10 wt% ZIF-zni@PU | 1.43 ± 0.03 |
| 15 wt% ZIF-zni@PU | 1.48 ± 0.03 |
| 20 wt% ZIF-zni@PU | 1.73 ± 0.01 |
| 25 wt% ZIF-zni@PU | 1.71 ± 0.02 |

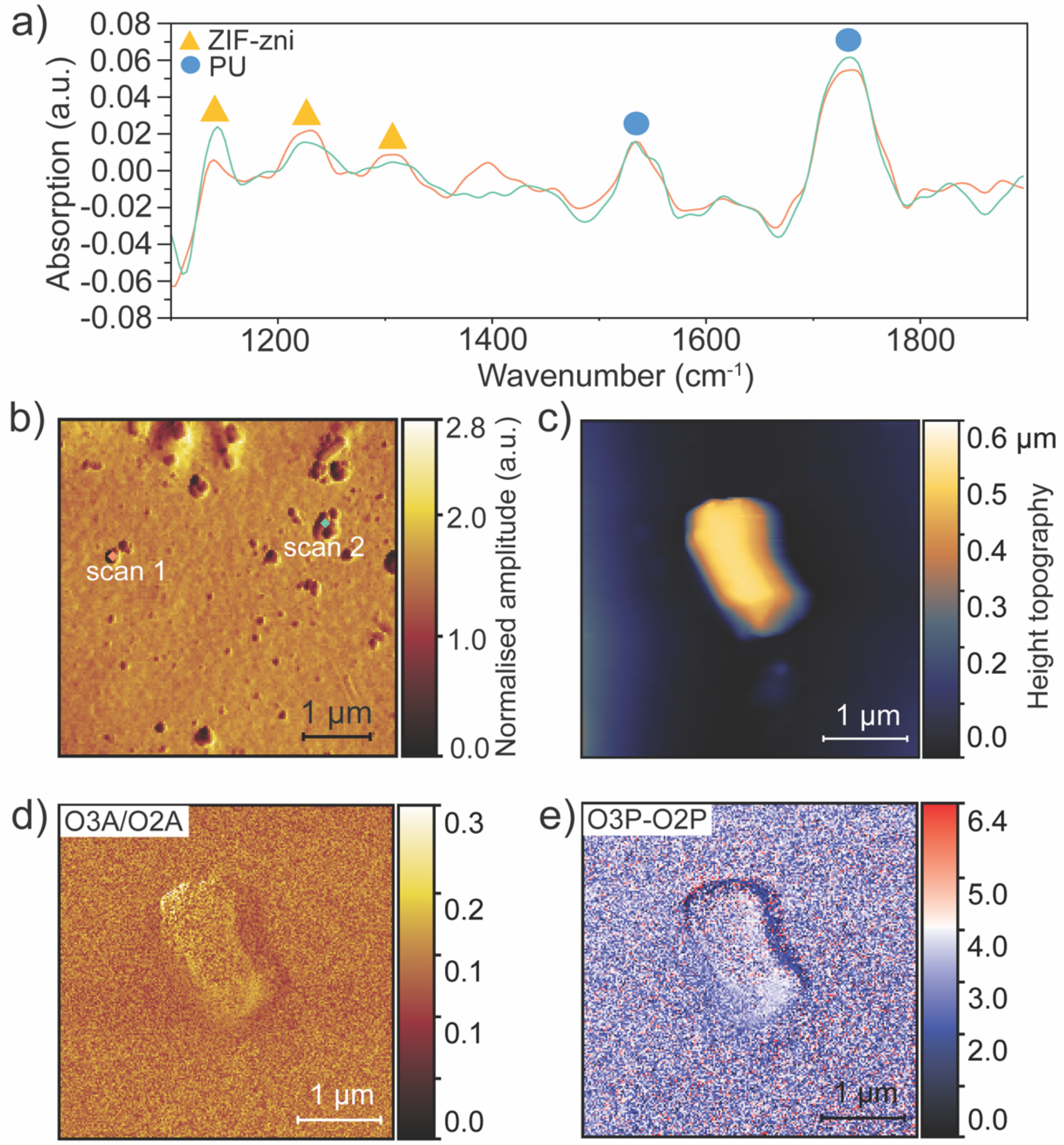

**Figure S7.** a) Nano-FTIR point spectra collected at two locations on the PDMS surface after extended overnight contact–separation cycles. The yellow and green curves represent Scan 1 and Scan 2, respectively. b) Normalised O2A image of the cycled PDMS surface with two selected points for spectral acquisition as presented in panel (a). c) Height topography of a 20 wt% ZIF-zni@PU composite film showing ZIF-zni crystals. PsHet signal normalization employing the consecutive optical harmonics at 1496 $cm^{-1}$, namely d) O3A/O2A and e) O3P−O2P, used to eliminate far-field scattering, reflection, and background contributions.

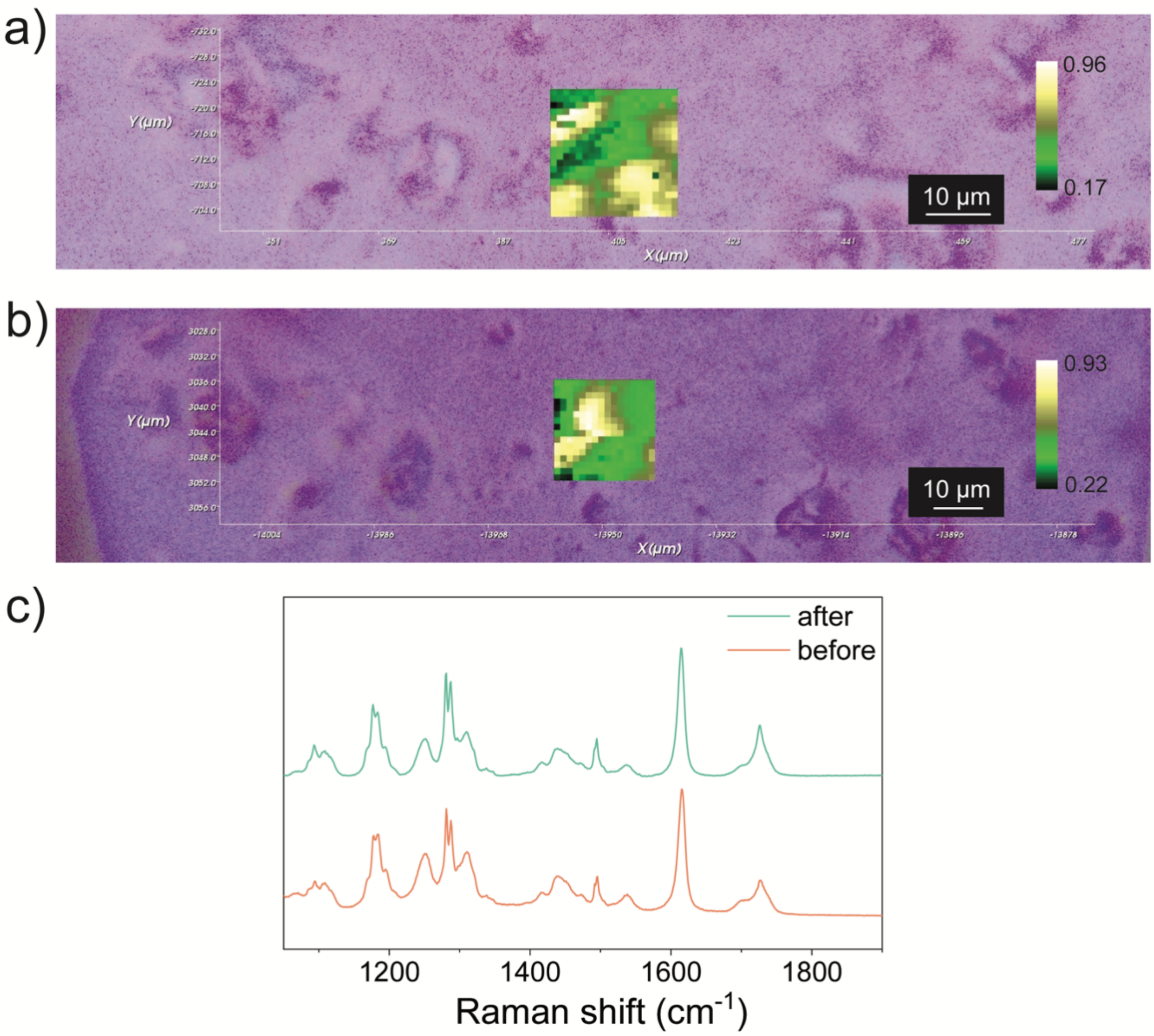

**Figure S8.** Raman spectral map a) before and b) after extended overnight contact-separation testing using a freshly prepared 20 wt% ZIF-zni@PU composite film against pristine PDMS. The same area was scanned before and after the extended TENG test. The average spectrum for each scanned area was baseline-corrected, normalized, vertically offset, and shown in c). Standard TENG testing conditions were applied during the overnight experiment.

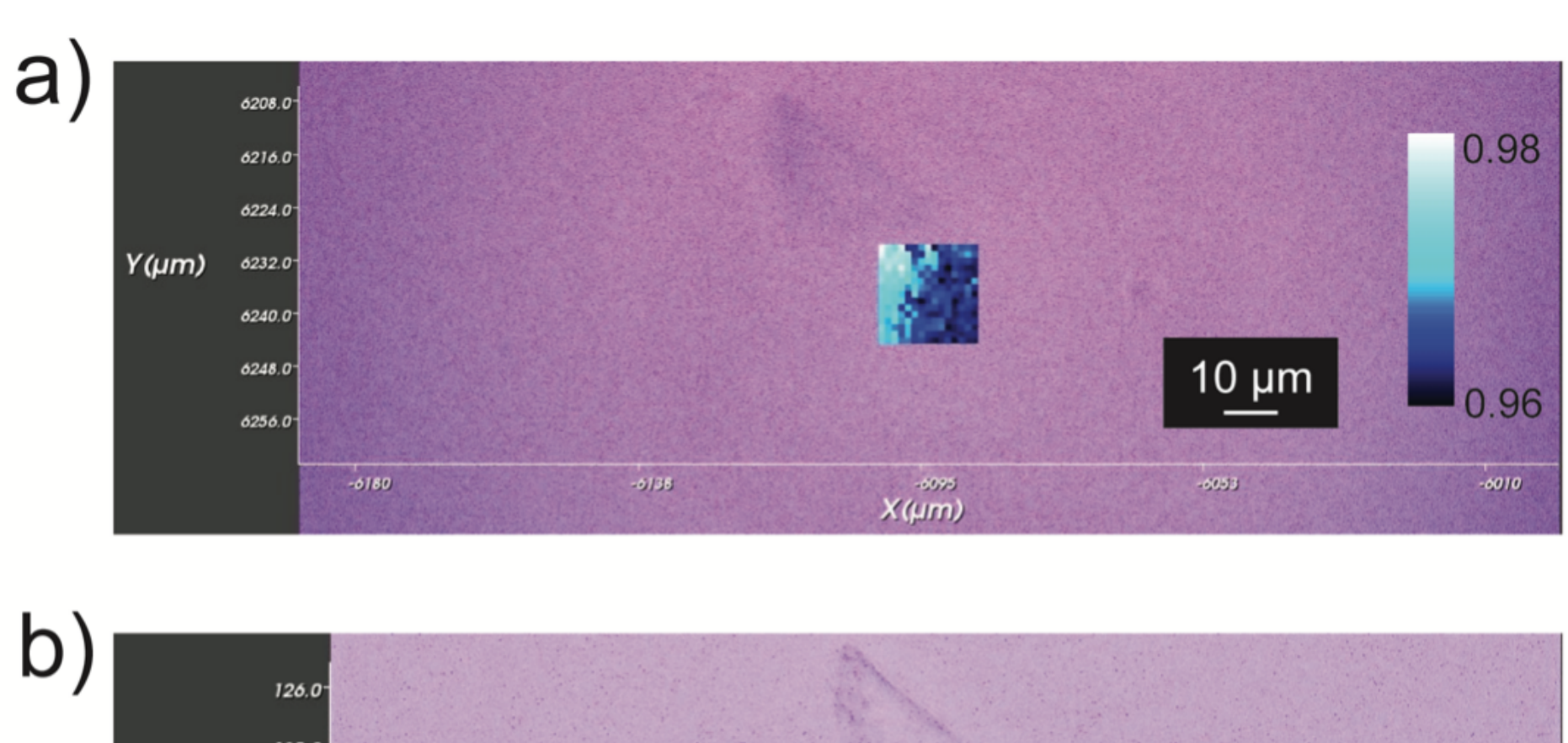

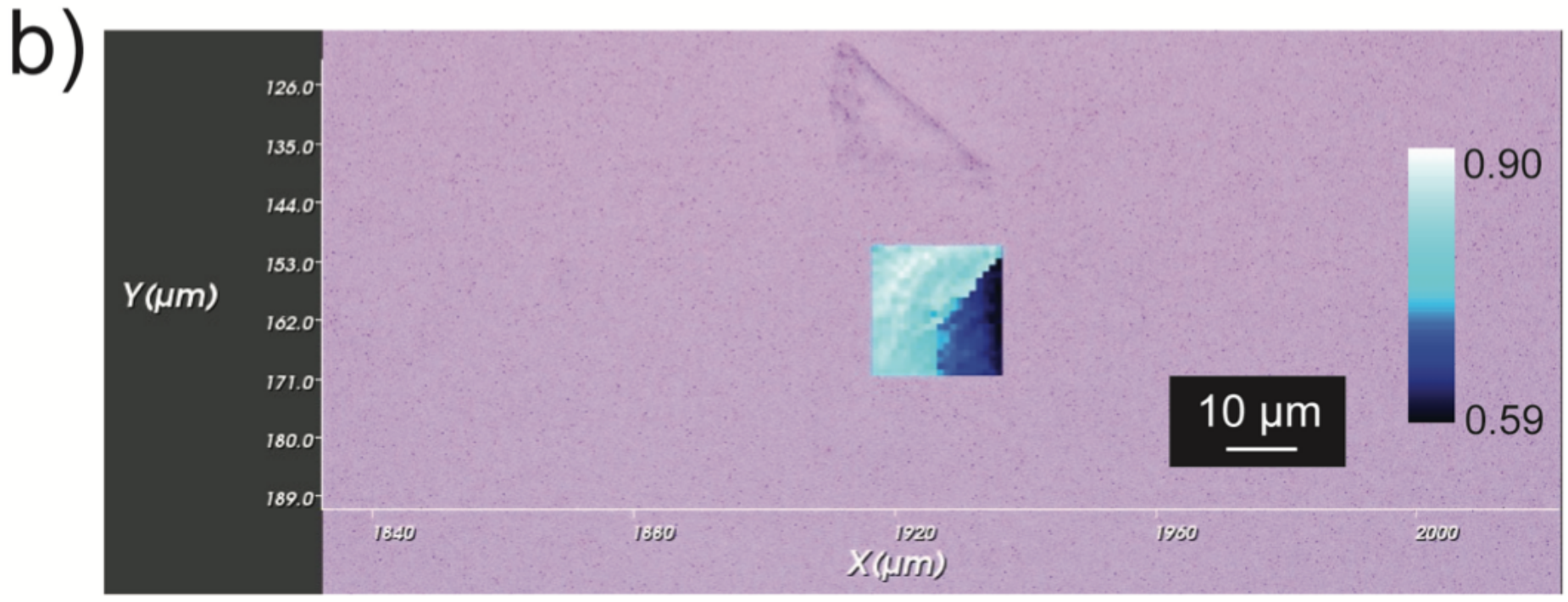

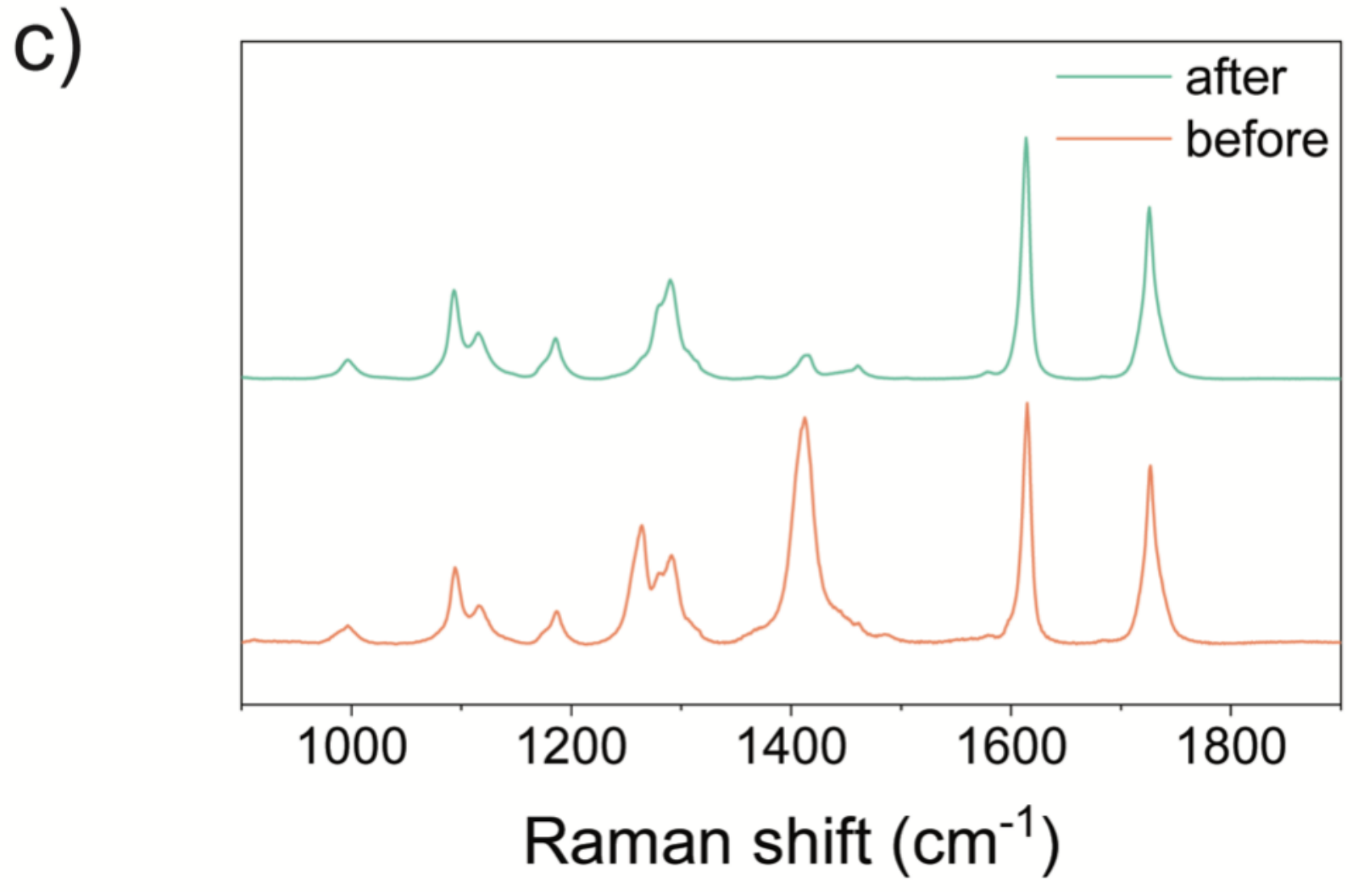

**Figure S9.** Raman spectra of PDMS tribonegative side a) before and b) after overnight contact-separation test. The scanned area was divided into two regions, light blue and dark blue. The average spectrum of each scanned area was baseline-corrected, normalized, vertically offset, and shown in c). The change in peak intensity can be attributed to viscoelastic strain induced by the impact test.

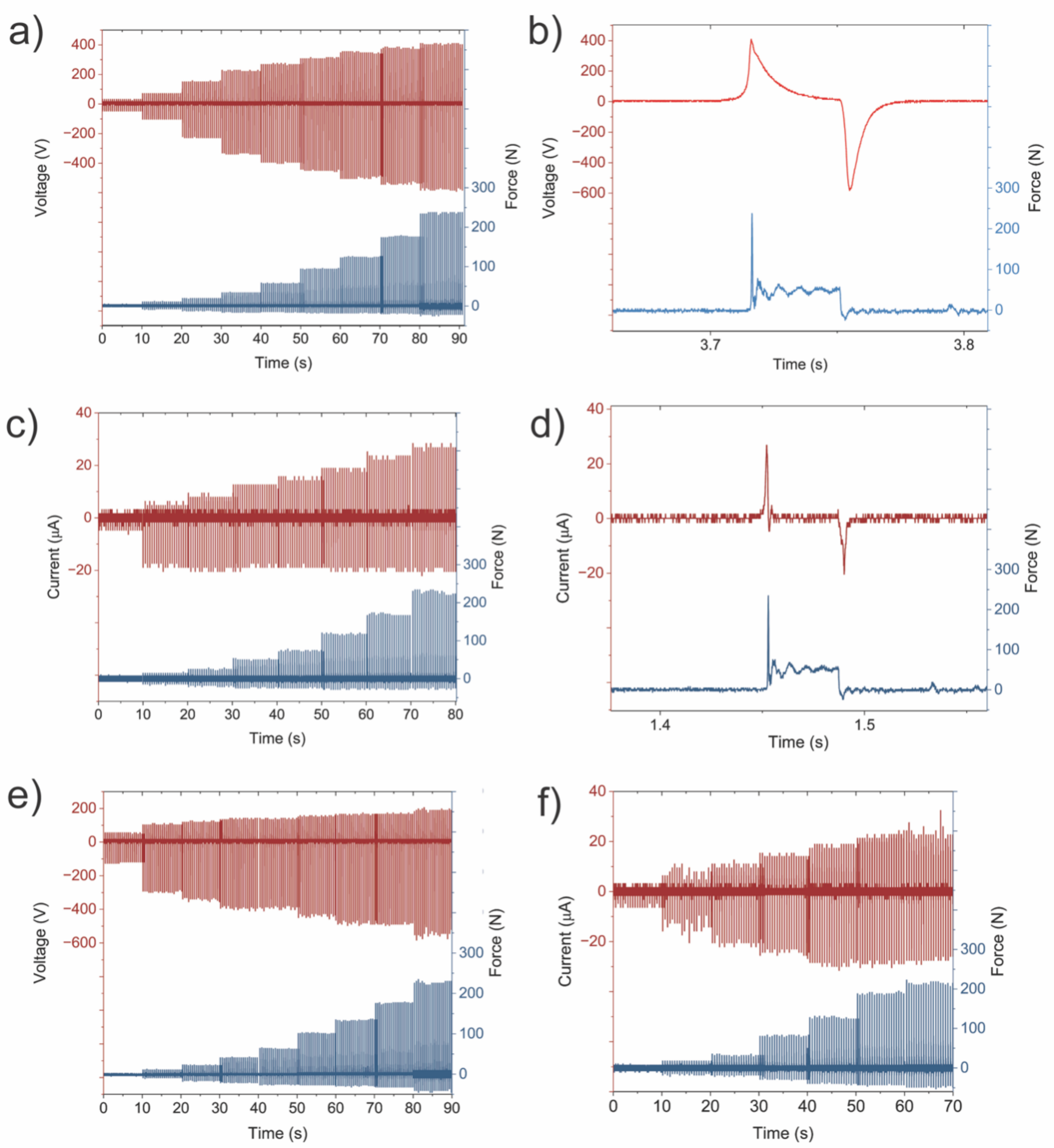

**Figure S10.** Comparison between the 20 wt% ZIF-zni@PU TENG vs. PU TENG using the high-load LDS V400 electromagnetic shaker. a) Voltage output with an increasing force (up to 230 N) for 20 wt% ZIF-zni@PU TENG. b) Single voltage peak correlated to ~230 N force using 20 wt% ZIF-zni@PU TENG. c) Current output with peak force for 20 wt% ZIF-zni@PU TENG; d) single peak with corresponding peak force ~230 N. e) Voltage output and f) current output with a similar increase of impact forces (up to 230 N) of PU TENG.

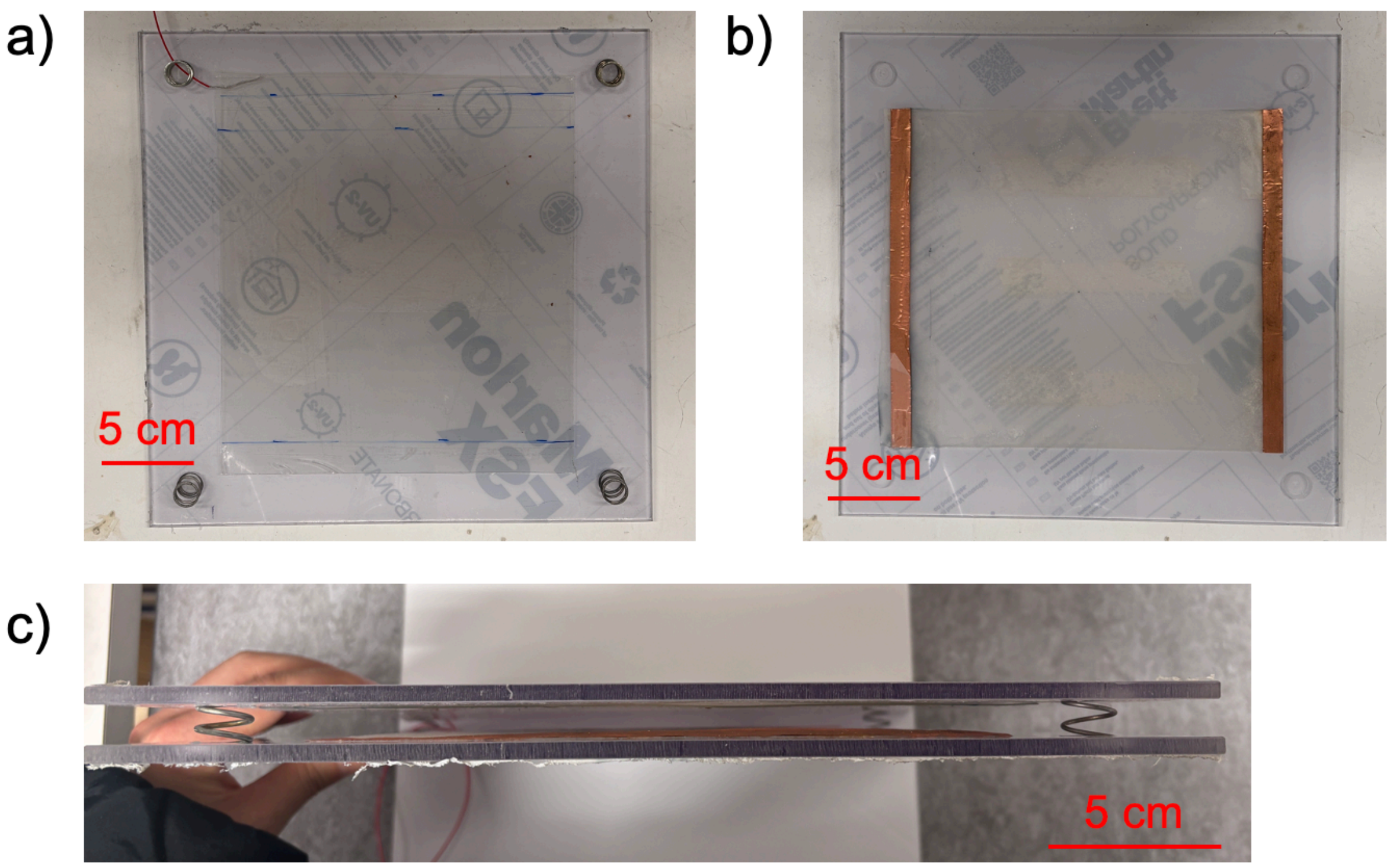

**Figure S11.** Structural assembly and device configuration of a 20 wt% ZIF-zni@PU vs PDMS based triboelectric floor tile prototype. a) PDMS film of 15 cm × 15 cm. b) 20 wt% ZIF-zni@PU composite film. c) Side view of floor prototype with a gap spacing of 1 cm and polycarbonate thickness of 0.5 cm.

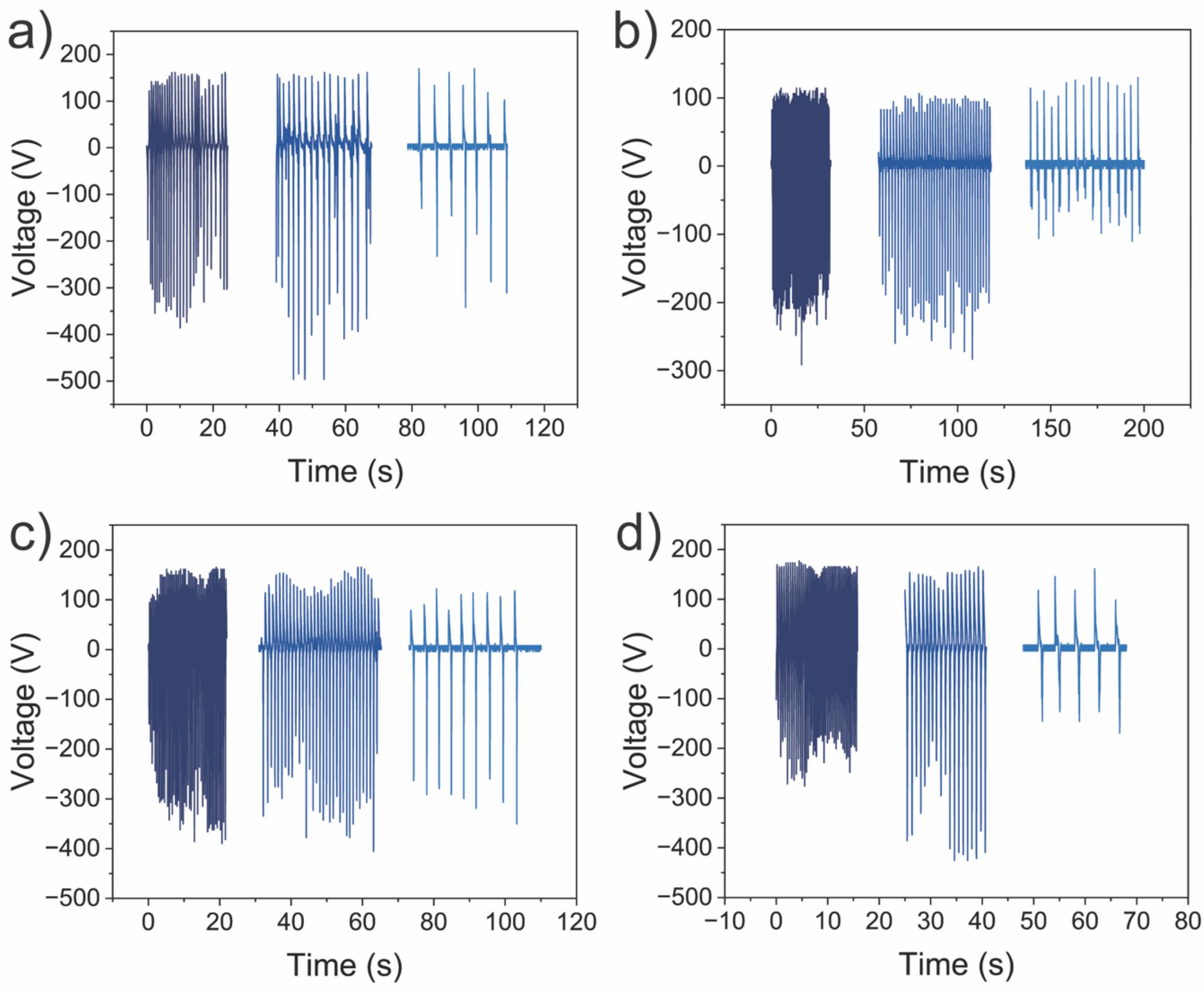

**Figure S12.** TENG floor-tile voltage outputs from four additional participants with body masses of a) 55 kg, b) 60 kg, c) 75 kg, and d) 80 kg, measured at fast, medium, and slow stepping rates (left to right).

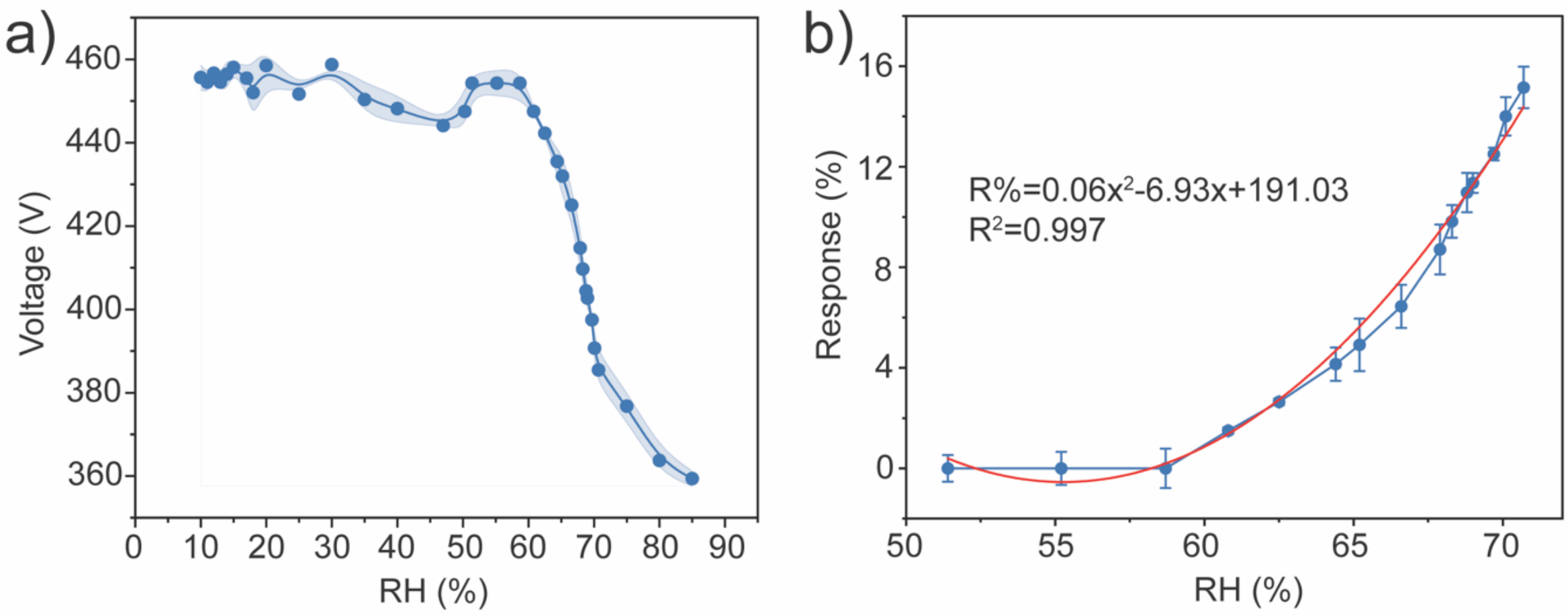

**Figure S13.** Humidity-responsive triboelectric behaviour of the 20 wt% ZIF-zni@PU TENG. a) Output voltage of the 20 wt% ZIF-zni@PU TENG as a function of relative humidity (RH) over the nominal range of 10–85% RH. b) Response (%) versus RH at room temperature. The red line represents a quadratic curve fit of the experimental data. Error bars indicate the standard deviation (*n* = 3).

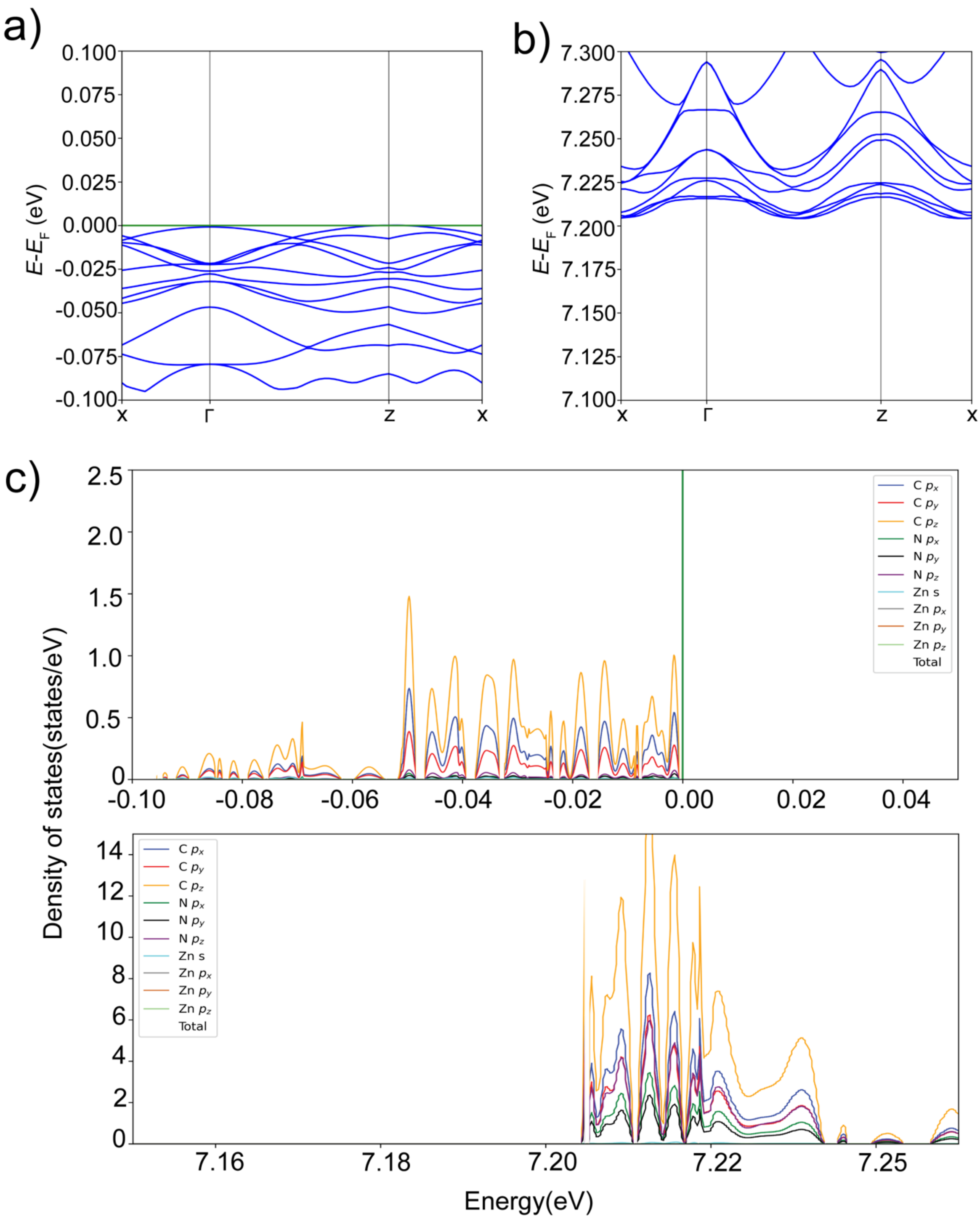

**Figure S14.** DFT calculated electronic band structure of ZIF-zni computed along the high-symmetry Brillouin-zone path **X–Γ–Z–X**. Owing to the exceptionally large band gap, the energy dispersion is presented separately for a) the valence bands and b) the conduction bands. The Fermi level is referenced to 0 eV. c) Projected density of states (PDOS) near the valence band (VB) and conduction band (CB), showing calculated VB contributions from delocalized imidazolate $\pi/\sigma$ states and localized CB contributions from antibonding $\sigma^*$ states. Note that these are calculated localization electronic structure trends, which are not direct evidence of interfacial charge transfer.

**Details of Computational Methods Employing DFT**

*Ab initio* density functional theory (DFT) and density functional perturbation theory (DFPT) calculations were performed to investigate the electronic, dielectric, and electromechanical properties of ZIF-zni. All periodic calculations were carried out using the CRYSTAL 23 code[1], which employs an all-electron, localized Gaussian-type orbital (GTO) basis set and treats crystalline orbitals as linear combinations of Bloch functions constructed from atom-centered basis functions. This approach enables accurate description of both electronic structure and lattice-dynamical properties in extended framework materials.

The exchange–correlation energy was described using the PBE0 hybrid functional[2], which incorporates a fixed fraction of exact Hartree-Fock exchange and is well suited for predicting band gaps and polarization properties in periodic framework solids, such as MOFs. Dispersion interactions were included *via* the Grimme D3 correction, ensuring an accurate treatment of long-range *van der Waals* forces relevant to dense framework packing. All-electron basis sets of double-$\zeta$ quality were employed for Zn, C, N, and H atoms, yielding a total of approximately $1.3 \times 10^4$ localized basis functions per unit cell.

Full geometry optimization of both lattice parameters and internal atomic coordinates was performed using a quasi-Newton algorithm until convergence thresholds of $10^{-5}$ a.u. for forces and $10^{-4}$ a.u. for atomic displacements were simultaneously satisfied. Reciprocal-space sampling was carried out using a Monkhorst-Pack shrinking factor of 2, and the truncation thresholds for bi-electronic integrals (TOLINTEG) were set to 7 7 7 7 25. Self-consistent field (SCF) convergence criteria were set to $10^{-7}$ Hartree for geometry optimizations and $10^{-10}$ Hartree for subsequent property calculations.

Vibrational frequencies and infrared (IR) intensities were computed at the $\Gamma$-point using DFPT within the coupled-perturbed Hartree-Fock/Kohn-Sham (CPHF/KS) formalism[3]. The dynamical matrix was obtained *via* numerical differentiation of analytical gradients, and IR intensities were evaluated using the Berry-phase polarization approach. Simulated IR spectra were generated by applying Lorentzian broadening with a full width at half maximum (FWHM) of 8-10 $cm^{-1}$, and a uniform scaling factor of 0.95 was applied to facilitate comparison with the experimental spectra[4].

Born effective charge tensors, frequency-dependent dielectric functions, and static dielectric constants were extracted directly from DFPT calculations. Real-space electrostatic potential (ESP) maps, projected density of states (PDOS), and highest occupied and lowest unoccupied crystalline orbitals (HOCO and LUCO) were computed at the same level of theory. All electronic structure visualizations were rendered using JMol and post-processed with customized Python scripts[3].

**Movie Clips**

**Movie S1:** Real-time testing of the 20 wt% ZIF-zni@PU TENG vs PDMS for illuminating 240 LEDs in the dark, under ambient conditions.

**Movie S2**: Example of two people stepping on the 20 wt% ZIF-zni@PU vs PDMS based triboelectric floor tile.

**Movie S3**: Animation of the calculated lattice vibrational mode of ZIF-zni at 27 $cm^{-1}$.

**Movie S4:** Animation of the calculated lattice vibrational mode of ZIF-zni at 50 $cm^{-1}$.

**Movie S5:** Animation of the calculated lattice vibrational mode of ZIF-zni at 80 $cm^{-1}$.

**Movie S6:** Animation of the calculated lattice vibrational mode of ZIF-zni at 350 $cm^{-1}$.

**Movie S7:** Animation of the calculated lattice vibrational mode of ZIF-zni at 650 $cm^{-1}$.

**References**

1. Alessandro Erba, J. K. D., Silvia Casassa, Bartolomeo Civalleri, Lorenzo Donà, Ian J. Bush, Barry Searle, Lorenzo Maschio, Loredana Edith-Daga, Alessandro Cossard, Chiara Ribaldone, Eleonora Ascrizzi, Naiara L. Marana, Jean-Pierre Flament, Bernard Kirtman. CRYSTAL23: A Program for Computational Solid State Physics and Chemistry. *J. Chem. Theory Comput.* **19**, 6891–6932 (2023). https://doi.org:10.1021/acs.jctc.2c00958
2. L. Doná, J. G. B., B. Civalleri. Extending and assessing composite electronic structure methods to the solid state. *J. Chem. Phys* **151** (2019). https://doi.org:10.1063/1.5123627
3. Hanson RM, L. X. DSSR-enhanced visualization of nucleic acid structures in Jmol. *Nucleic Acids Res.* **45**, W528-W533 (2017). https://doi.org:10.1093/nar/gkx365